\documentclass[12pt]{article}%
\usepackage{amsfonts}
\usepackage{amsmath,amssymb}
\usepackage{graphicx}%
\usepackage{amsmath}%
\pdfoutput=1
\makeatletter

\@addtoreset{equation}{section}
\makeatother
\newcommand{\beq}{\begin{equation}}
\newcommand{\eeq}{\end{equation}}
\renewcommand{\a}{\alpha}

\begin{document}

\baselineskip=18pt
%a la harvmac
\baselineskip 0.7cm

\begin{titlepage}
%% Set the number of the title with 0
\setcounter{page}{0}
% change the footnote symbol
\renewcommand{\thefootnote}{\fnsymbol{footnote}}
%------------------
\begin{flushright}
%preprint number
%CALT-mm-nnnn\\
%IPMU09-nnnn\\
%UT-09-30\\
%month, 2008
\end{flushright}
%---------------------
\vskip 1.5cm
\begin{center}
{\LARGE \bf
Supergravity Actions  with
Integral Forms
\vskip 1.5cm
}
{\large
L. Castellani$^{~a,b,}$\footnote{leonardo.castellani@mfn.unipmn.it},
R. Catenacci$^{~a,}$\footnote{roberto.catenacci@mfn.unipmn.it},
and
P.A. Grassi$^{~a,b,}$\footnote{pietro.grassi@mfn.unipmn.it}
\medskip
}
\vskip 0.5cm
{
\small\it
\centerline{$^{(a)}$ Dipartimento di Scienze e Innovazione Tecnologica, Universit\`a del Piemonte Orientale} }
\centerline{\it Viale T. Michel, 11, 15121 Alessandria, Italy}
\medskip
\centerline{$^{(b)}$ {\it
INFN, Sezione di Torino, via P. Giuria 1, 10125 Torino} }
\vskip  .5cm
\medskip
\end{center}
%-----------------------------------------
\centerline{{\bf Abstract}}
\medskip
\noindent
Integral forms provide a natural and powerful tool for the construction of
supergravity actions. They are generalizations of usual
differential forms and are needed for a consistent theory of integration
on supermanifolds. The group geometrical approach to supergravity
and its variational principle
are reformulated and clarified in this language.
Central in our analysis is the Poincar\'e dual of a bosonic manifold
embedded into a supermanifold. Finally, using integral forms we provide a proof of
Gates' so-called ``Ectoplasmic Integration Theorem'', relating superfield actions to component actions.
\bigskip
\bigskip

\noindent
\today
\end{titlepage}
\setcounter{page}{1}
%don't number title page

\vfill
\eject

\tableofcontents
\vfill
\eject
\newpage\setcounter{footnote}{0} \newpage\setcounter{footnote}{0}
%%%%%%%%%%%%%%%%%%%%%%%%%%%%%%%%%%%%%%%%%%%%%%%%%%%%%%%%%%%%%%%%%%%%%%%

\section{Introduction}

In the study of quantum field theories, of string theory and several other
modern theoretical models the action is a fundamental bookkeeping device for
all needed constraints, equations of motion and quantum corrections. In many
cases having an action has tremendous advantages over the only knowledge of
the equations of motion or other auxiliary constraints. In particular, the
action encodes both the dynamics of the theory and the symmetries of the model
(by means of Noether theorem) in a very compact formulation. Nonetheless,
there are several situations where the construction of an action does not seem
possible or out of the reach by present means. For example, it is not known
whether a manifestly supersymmetric $N=4, D=4$ super-Yang-Mills action exists
in superspace (which would guarantee the well-known renormalisation theorems),
and this is due to the self-duality constraints and to the lacking of an
off-shell superspace formulation. Again, no standard superspace action for
type IIB $D=10$ supergravity theory exists, due to the self-duality
constraints on RR fields. For the same reasons, no superspace formulation of
$N=2, D=6$ supergravity is known.

Furthermore, even when the superspace formulation exists, it is difficult to
extract the component action. This happens mainly for supergravity theories,
where the superdeterminant of the supervielbein is needed for the construction
of the action. In many cases, that computation is very cumbersome. On the
other side in the work of Gates et al.
\cite{Gates:1997kr,Gates:1997ag,Gates:1998hy,Gates:2009uv} a new method is been provided to
extract the component action from the superspace formulation. This is based on
a formula which relates the superfield action to the component action via a
density projection operator acting on a closed superform. This procedure
incorporates the integration over the fermionic coordinates and the contributions
due to the gravitons. We show here that the origin of
that formula can be understood by interpreting the superfield action  as an
integral form. The relation between the density projection operator and the
component action is achieved by partial integration using picture changing operators.

Three decades ago a group-based geometric approach to supergravity was put
forward, known as group manifold approach \cite{Castellani,Castellani:1981},
intermediate between the superfield and the component approaches. This
framework provides a systematic algorithm to construct supergravities in any
dimension. The starting point is a supergroup, and the fields of the theory
are identified with the vielbein one-forms of (a manifold diffeomorphic to)
the supergroup manifold. For example in D=4, N=1 supergravity the dynamical
fields are the vierbein, the spin connection and the gravitino one-forms, dual
respectively to the translation, Lorentz rotations and supersymmetry tangent
vectors. Thus supermultiplets come out of supergroups, rather than from a
superfield depending on bosonic and fermionic coordinates. Actions in $D$
dimensions are constructed by considering integrals of $D$-form Lagrangians
$L$ on $D$-dimensional submanifolds of the supergroup manifold. The action
depends in general also on how the submanifold is chosen inside the supergroup
manifold, and the action principle includes also variations in the submanifold
embedding functions. The resulting field equations are $(D-1)$-form equations
holding on the whole supergroup manifold. The way to relate these actions and
their field equations to those of the ``ordinary" $D$-dimensional
supergravities is exhaustively illustrated by many examples in ref.
\cite{Castellani}. One of the advantages of this approach is that it yields
the self-duality constraints of the $D=6$ and $D=10$ supergravities mentioned
above as \textit{part of the equations of motion}, besides allowing to
construct the corresponding actions \cite{D'Auria:1983,Castellani:1987}.

We show here how the variational principle of the group manifold approach can
be reformulated and clarified by using integral forms and the Poincar\'e dual
of the submanifold. In particular we derive the condition for the embedding
independence of the submanifold. This coincides with the condition for local
supersymmetry invariance of the spacetime action, and reduces to the vanishing
of the contraction of $dL$ along tangent vectors orthogonal to the submanifold.

The paper has the following organisation. In sec. 2, the integration on
supermanifolds is briefly discussed and presented both from a mathematical
point of view, and from a more intuitive/physical point of view. The
integration on curved supermanifolds is also discussed. In sec. 3, we
describe, also for the case of supermanifolds, a simple and explicit form  of
the Poincar\'e dual as a singular localization form. The integration on a
submanifold and the independence of the embedding is discussed. The
construction of the actions in the group-geometric approach is presented and
the variational principle is explained. Finally, in sec. 4 we consider the
relation between the integral of superforms in the ectoplasmic integration
formalism and integral forms. The ``ethereal conjecture" of Gates et al.
\cite{Gates:1997kr,Gates:1997ag,Gates:1998hy,Gates:2009uv} is proved using integral forms.
Appendix A contains some additional material ancillary to the main text.

%%%%%%%%%%%%%%%%%%%%%%%%%%%%%%%%%%%%%%%%%%%%%%%%%%%%%%%%%%%%%%%%%%%%%%%%%

\section{Integration on Supermanifolds}

In this section we give a short introduction to the theory of integration on
supermanifolds (see for example the review by Witten \cite{Witten:2012bg}). The
translation of the {\it{picture changing operators}} into supergeometry has been explored
in \cite{integ,Voronov2}. More recently, the application to
target space supersymmetry and Chern-Simons theories have been discussed in \cite{Grassi:2004tv}. Thom classes for supermanifolds
have been constructed in \cite{mare}. The picture changing operators have been introduced in string theory in \cite{FMS},
from world sheet point of view, and in
\cite{Berkovits:2004px}, from target space point of view.

We start, as usual, from the case of the real superspace $\mathbb{R}^{n|m}$
with $n$ bosonic ($x^{i},i=1,\dots,n$) and $m$ fermionic $(\theta^{\alpha
},\alpha=1,\dots,m$) coordinates. We take a function $f(x,\theta)$ in
$\mathbb{R}^{n|m}$ with values in the real algebra generated by $1$ and by the
anticommuting variables, and we expand $f$ as a polynomial in the variables
$\theta:$%
\[
f(x,\theta)=f_{0}(x)+...+f_{m}(x)\theta^{1}...\theta^{m}%
\]
If the real function $f_{m}(x)$ is integrable in some sense in $\mathbb{R}%
^{n},$ the Berezin integral of $f(x,\theta)$ is defined as:%
\[
\int_{\mathbb{R}^{n|m}}f(x,\theta)[d^{n}xd^{m}\theta]=\int_{\mathbb{R}^{n}%
}f_{m}(x)d^{n}x
\]
Note that $d^{n}x \equiv dx^{1}\wedge...\wedge dx^{n}$ is a volume form (a top
form) in $\mathbb{R}^{n}$, but $[d^{n}x\,d^{m}\theta]$ is just a formal symbol
that has nothing to do neither with \textquotedblleft exterior products", nor
with \textquotedblleft top forms" mainly because if $\theta$ is a fermionic
quantity, $d\theta$ is bosonic $(d\theta\wedge d\theta\neq0).$

An important property of $[d^{n}x\,d^{m}\theta]$ is elucidated by the
following simple example: consider in $\mathbb{R}^{1|1}$ the function
$f(x,\theta)=g(x)\theta$ (with $g(x)$ integrable function in $\mathbb{R}$ ).
We have:%
\[
\int_{\mathbb{R}^{1|1}}f(x,\theta)[dxd\theta]=\int_{\mathbb{-\infty}}%
^{+\infty}g(x)dx\,.
\]
If we rescale $\theta\rightarrow\lambda\theta$ ( $\lambda\in\mathbb{R}$ ) we
find $f(x,\theta)\rightarrow\lambda f(x,\theta)$. For the integral to be
invariant under coordinate changes, the ``measure" $[dxd\theta]$ must rescale
as $[dxd\theta]\rightarrow\frac{1}{\lambda}[dxd\theta]$ and not as
$[dxd\theta]\rightarrow\lambda\lbrack dxd\theta]$.

Generalizing this fact it is known that under general coordinate
transformations in superspace the symbol $[d^{n} x \, d^{m}\theta]$ transforms
with the ``Berezinian", a.k.a. the superdeterminant, while $d^{n}x $
transforms in $\mathbb{R}^{n}$ with the Jacobian determinant. This fact is
very important, because supermanifolds are obtained by gluing together open
sets\footnote{The most natural topology in $\mathbb{R}^{n|m}$ is the topology
in which the open sets are the \textbf{complete} cylinders over open sets in
$\mathbb{R}^{n}.$ This ``coarse" topology is then transferred to the
supermanifold.} homeomorphic to $\mathbb{R}^{n|m}.$ The transformation
properties (i.e. transition functions) allow to define integration on
supermanifolds. The concept of \textbf{integral forms }arises also for giving
a definite meaning to the symbols $[d^{n}xd^{m}\theta]$ by specifying what
kind of object is "integrated".

A brief review of the formal properties of integral forms
\cite{integ,Grassi:2004tv} is given in Appendix A. Here we elaborate on their
definition and on the computation of integrals.

The usual integration theory of differential forms for bosonic manifolds can
be conveniently rephrased to shed light on its relations with Berezin integration.

We start again with a simple example: consider in $\mathbb{R}$ the integrable
1-form $\omega$ $=g(x)dx$ (with $g(x)$ integrable function in $\mathbb{R}$ ).
We have:
\[
\int_{\mathbb{R}}\omega=\int_{\mathbb{-\infty}}^{+\infty}g(x)dx\,.
\]
Observing that $dx$ is an anticommuting quantity, and denoting it by $\psi$,
we could think of $\omega$ as a function on $\mathbb{R}^{1|1}$:%
\begin{equation}
\omega=g(x)dx=f(x,\psi)=g(x)\psi
\end{equation}
This function can be integrated \textit{\`{a} la} Berezin reproducing the
usual definition:%
\[
\int_{\mathbb{R}^{1|1}}f(x,\psi)[dxd\psi]=\int_{\mathbb{-\infty}}^{+\infty
}g(x)dx\,=\int_{\mathbb{R}}\omega
\]
Note that (as above) the symbol $[dxd\psi]$ is written so as to emphasize that
we are integrating on the \textbf{two} variables $x$ and $\psi$, hence the
$dx$ inside $[dx d\psi]$ is \textit{not} identified with $\psi$.

This can be generalized as follows. Denoting by $M$, a bosonic differentiable
manifold with dimension $n$, we define the exterior bundle $\Omega^{\bullet
}(M)=\sum_{p=0}^{n}\bigwedge^{p}(M)$  as the direct sum of $\bigwedge^{p}(M)$
(sometimes denoted also by $\Omega^{p}(M)$). A section $\omega$ of
$\Omega^{\bullet}(M)$ can be written locally as
\begin{equation}
\omega=\sum_{p=0}^{n}\omega_{\lbrack i_{1}\dots i_{p}]}(x)dx^{i_{1}}%
\wedge\dots\wedge dx^{i_{p}}\label{inA}%
\end{equation}
where the coefficients $\omega_{\lbrack i_{1}\dots i_{p}]}(x)$ are functions
on $M$ and the indices $[i_{1},\dots,i_{p}]$ are antisymmetrized. The integral
of $\omega$ is defined as:
\begin{equation}
I[\omega]=\int_{M}\omega=\int_{M}\epsilon^{i_{1}\dots i_{n}}\omega_{\lbrack
i_{1}\dots i_{n}]}(x)\,d^{n}x\,,\label{inB}%
\end{equation}
At first sight this might seem a bit strange, but we are actually saying that
in the definition of the integral only the ``part of top degree" of $\omega$
is involved. This opens the way to the relations between the integration
theory of forms and the Berezin integral, that can be exploited by
substituting every $1$-form $dx^{i}$ with a corresponding Grassmann variable
$\psi^{i}.$ A section $\omega$ of ${\Omega^{\bullet}(}M{)}$ is viewed locally as a
function on a supermanifold $\mathcal{M}$ with coordinates $(x^{i},\psi^{i})$
\begin{equation}
\omega(x,\psi)=\sum_{p=0}^{n}\omega_{\lbrack i_{1}\dots i_{p}]}(x)\psi^{i_{1}%
}\dots\psi^{i_{p}}\,;\label{inAA}%
\end{equation}
such functions are polynomials in $\psi$'s. Supposing now that the form
$\omega$ is integrable we have as above that the Berezin integral ``selects"
the top degree component of the form:%
\begin{equation}
\int_{\mathcal{M}}\omega(x,\psi
)[d^{n}xd^{n}\psi]=\int_{M}\omega
\end{equation}

If the manifold is equipped with a metric $g$ (that for the moment we assume
globally defined), we can expand a generic form $\omega$ on the basis of forms
$\psi^{a} = e_{i}^{a}dx^{i}$ ($a=1,\dots,n$) such that $g=\psi^{a}\otimes
\psi^{b}\eta_{ab}$ where $\eta^{ab}$ is the flat metric on the tangent space
$T(M)$ and we have that
\begin{equation}
I[\omega,g]=\int_{\mathcal{M}}\omega(x,e)[d^{n}xd^{m}\psi]=\int_{M}%
e\,\epsilon^{i_{1}\dots i_{n}}\omega_{\lbrack i_{1}\dots i_{n}]}(x)d^{n}%
x=\int_{M}\sqrt{g}\,\epsilon^{i_{1}\dots i_{n}}\omega_{[i_{1}\dots i_{n}%
]}(x)\,d^{n}x\,,\label{inC}%
\end{equation}
where $e={\det}(e_{i}^{a})$, and $g=\det(g_{ij})$. Again, we use the Berezin
integral to select the top degree component of the form. Notice that the last
integral can be computed if suitable convergence conditions are satisfied
according to Riemann or Lebesgue integration theory.
%%For example if the manifold ${\cal M}$ is not compact and if $\sqrt{g}$ has a suitable asymptotical behavior,
%%a large class of function can be integrated. On the other side, if ${\cal M}$ is compact and integrand has integrable
%%singularities, we have a meaningful expression.

In the following we will need also distributions, and therefore we consider
expressions that factorize $\epsilon^{i_{1}\dots i_{n}}\omega_{\lbrack
i_{1}\dots i_{n}]}(x)$ into a distributional part $\frac{1}{\sqrt{g}}%
\prod_{i=1}^{n}\delta(x^{i})$ (the additional $\frac{1}{\sqrt{g}}$ is added
for covariance under diffeomorphisms) and into a test function $\tilde{\omega
}(x)$ (for example, belonging to the space of fast decreasing functions). In
this case:
\begin{equation}
\int_{M}\omega=\int_{M}\tilde{\omega}(x)\left[  \prod_{i=1}^{n}\delta
(x^{i})\right]  d^{n}x=\tilde{\omega}(0)\,,\label{inD}%
\end{equation}
where in the last term we evaluate the expression at $x^{i}=0$. In this case,
the compactness of the space or other convergence conditions do not matter,
since the measure is concentrated in the point $x^{i}=0$. The points $x_{i}$
where the integral is localised can be moved by suitable diffeomorphisms.

%In QFT, we are used to expressions like
%\begin{equation}\label{inE}
%\int_{T^\bullet{\cal M}}  F \wedge F =  \int_{\cal M}  \e^{ijkl} F_{ij} F_{kl}
%\end{equation}
%where $n=4$, $F$ is  a 2-form. The advantage of having a form integral like as in the left hand side is
%the independence upon the coordinate system and explicit covariance under diffeomorphisms of the manifold
%${\cal M}$. In addition, we can use the Stokes theorem: if $\omega = d \eta$ then we
%have
%\begin{equation}\label{inF}
%\int_{T^\bullet{\cal M}} d \eta(x, e)  = \int_{ \partial T^\bullet{\cal M}} \eta(x, e)
%\end{equation}
%where ${\partial T^\bullet{\cal M}}$ is the exterior bundle of the boundary manifold $\partial {\cal M}$.
%In order to built interesting Lagrangians, we need to construct top forms from low dimensional forms by
%wedge multiplications of forms according to the usual map
%\begin{equation}\label{inFA}
%\wedge: \wedge^{p} {\cal M} \times \wedge^{q} {\cal M} \longrightarrow \wedge^{p+q} {\cal M}\,.
%\end{equation}
%and by introducing the Hodge dual
%\begin{equation}\label{inFB}
%\star: \wedge^{p} {\cal M} \longrightarrow \wedge^{n-p} {\cal M}\,.
%\end{equation}

We denote now by $\mathcal{M}$ a supermanifold with coordinates $(x^{i}%
,\theta^{\alpha})$ (with $i=1,\dots,n$ and $\alpha=1,\dots,m$) and we consider
the ``exterior" bundle ${\Omega}^{\bullet}(\mathcal{M})$ as the direct sum of
bundles of fixed degree forms. The local coordinates in the total space of
this bundle are $(x^{i},d\theta^{\alpha},dx^{j},\theta^{\beta}),$ where
$\left(  x^{i},d\theta^{\alpha}\right)  $ are bosonic and $\left(
dx^{j},\theta^{\beta}\right)  $ fermionic. In contrast to the pure bosonic
case, a top form does not exist because the $1-$ forms of the type
$d\theta^{\alpha}$ commute among themselves $d\theta^{\alpha}\wedge
d\theta^{\beta}=d\theta^{\beta}\wedge d\theta^{\alpha}$. Then we can consider
forms of any degree
%(we will always omit from now on the wedge product symbol):
(wedge products are omitted in the following)
\begin{equation}
\omega=\sum_{p=0}^{n}\sum_{l=0}^{\infty}\omega_{\lbrack i_{1}\dots
i_{p}](\alpha_{1}\dots\alpha_{l})}(x,\theta)dx^{i_{1}}...dx^{i_{p}}%
d\theta^{\alpha_{1}}\dots d\theta^{\alpha_{l}}\label{inG}%
\end{equation}
where the coefficients $\omega_{\lbrack i_{1}\dots i_{p}](\alpha_{1}%
\dots\alpha_{l})}(x,\theta)$ are functions on the supermanifold $\mathcal{M}$
with the first $1\dots p$ indices antisymmetrized and the last $1\dots l$ symmetrized.

The component functions $\omega_{\lbrack i_{1}\dots i_{p}](\alpha_{1}%
\dots\alpha_{l})}(x,\theta)$ are polynomial expressions in the $\theta
^{\alpha}$ and their coefficients are functions of $x^{i}$ only. However, we
can adopt a different point of view: instead of simply expanding formally a
generic form $\omega(x,\theta,dx,d\theta)$ in $d\theta$, we can consider
analytic functions of the bosonic variables $d\theta$ and in addition we will
admit also distributions acting on the space of test functions of $d\theta$.
In this way, the exterior bundle in the $d\theta$ directions is a conventional
bosonic manifold with coordinates $d\theta^{\alpha}$ and the superforms become
distribution-valued on that space. In particular, we introduce the
distributions $\delta(d\theta^{\alpha})$ that have most (but not all!) of the
usual properties of the Dirac delta function $\delta(x)$. As explained in
Appendix A, one must have:
\begin{equation}
\delta(d\theta^{\alpha})\delta(d\theta^{\beta})=-\delta(d\theta^{\beta}%
)\delta(d\theta^{\alpha})\label{inH}%
\end{equation}
Therefore, the product of all Dirac's delta functions $\delta^{m}%
(d\theta)\equiv\prod_{\alpha=1}^{m}\delta(d\theta^{\alpha})$ serves as a
\textquotedblleft top form".

An \textbf{integral} form $\omega^{(p|q)}$ belonging to $\Omega^{(p|q)}%
(\mathcal{M})$ is characterised by two indices $(p|q)$: the first index is the
usual form degree and the second one is the \textit{picture} number which
counts the number of delta's. For a top form, that number must be equal to the
fermionic dimension of the space. Consequently, an \textit{integral} form
reads:
\begin{equation}
{\omega^{(p|q)}=\sum_{r=1}^{p}\omega_{\lbrack i_{1}\dots i_{r}](\alpha
_{r+1}\dots\alpha_{p})[\beta_{1}\dots\beta_{q}]}dx^{i_{1}}\dots{}dx^{i_{r}%
}d\theta^{\alpha_{r+1}}\dots{}d\theta^{\alpha_{p}}\delta(d\theta^{\beta_{1}%
})\dots\delta(d\theta^{\beta_{q}})}\label{integralform}%
\end{equation}
with $\omega_{\lbrack i_{1}\dots i_{r}](\alpha_{r+1}\dots\alpha_{p})[\beta
_{1}\dots\beta_{q}]}(x,\theta)$ superfields.

The $d\theta^{\alpha}$ appearing in the product and those appearing in the
delta functions are reorganised respecting the rule $d\theta^{\alpha}%
\delta(d\theta^{\beta})=0$ if $\alpha=\beta$. We see that if the number of
delta's is equal to the fermionic dimension of the space no $d\theta$ can
appear; if moreover the number of the $dx$ is equal to the bosonic dimension
the form $( $of type ${\omega^{(n|m)})}$ is a \textbf{top form.}

Notice that ${\omega^{(p|q)}}$ as written above is not the most generic form,
since we could have added the derivatives of delta functions (and they indeed
turn out to be unavoidable and will play an important role). They act by
reducing the form degree according to the rule $d\theta^{\alpha}\delta
^{\prime}(d\theta^{\alpha})=-\delta(d\theta^{\alpha})$, where $\delta^{\prime
}(x)$ is the first derivative of the delta function with respect to its
variable. (We denote also by $\delta^{(p)}(x)$ the p-derivative).

%where $\delta^{\prime}(d\theta^{a})%
%\equiv\iota_{a}\delta(d\theta^{a})$ (where $\iota_{a}$ is the contraction
%with respect to the supervector field $\partial_{a}$).

We also define as a \textit{superform} a $0$-picture integral form
$\Omega^{(p|0)}$
\begin{equation}
{\omega^{(p|0)}=\sum_{r=1}^{p}\omega_{\lbrack i_{1}\dots i_{r}](\alpha
_{r+1}\dots\alpha_{p})}dx^{i_{1}}\dots dx^{i_{r}}d\theta^{\alpha_{r+1}}\dots
{}d\theta^{\a_{p}}}\label{integralformII}%
\end{equation}%
\[
=\omega_{M_{1}\dots M_{p}}(Z)dZ^{M_{1}}\dots dZ^{M_{p}}%
\]
where the first indices are antisymmetrized while the spinorial indices
$\alpha_{1}\dots\alpha_{s}$ are symmetrized. In the last line, we have
collectively denoted by $Z^{M}$ the superspace coordinates and the indices
$M_{1}\dots M_{p}$ of the superform are graded symmetric.

Integral top forms (with maximal form degree in the bosonic variables and
maximal number of delta forms) are \textbf{the only objects we can hope to
integrate} on supermanifolds.

In general, if $\omega$ is a form in $\Omega^{\bullet}(\mathcal{M})$, its
integral on the supermanifold is defined as follows: (in analogy with the
Berezin integral for bosonic forms):
\begin{equation}
\int_{\mathcal{M}}\omega
\equiv\int_{\mathcal{M}}\epsilon^{i_{1}\dots i_{n}}\epsilon^{\beta_{1}%
\dots\beta_{m}}\omega_{\lbrack i_{1}\dots i_{n}][\beta_{1}\dots\beta_{m}%
]}(x,\theta)[d^{n}x\,d^{m}\theta]\label{inLAA}%
\end{equation}
where the last integral over $\mathcal{M}$ is the usual Riemann-Lebesgue
integral over the coordinates $x^{i}$ (if it is exists) and the Berezin
integral over the coordinates $\theta^{\alpha}$. The expressions
$\omega_{\lbrack i_{1}\dots i_{n}][\beta_{1}\dots\beta_{m}]}(x,\theta)$ denote
those components of (\ref{integralform}) with no symmetric indices.

Note that under the rescaling $\theta\rightarrow\lambda\theta$ ( $\lambda
\in\mathbb{R}$ ) the measure $[dx\,d\theta d\left(  dx\,\right)  d\left(
d\theta\right)  ]$ is an \textbf{invariant} quantity, in fact it is locally a
``product measure", and we know that $[dxd\theta]\rightarrow\frac{1}{\lambda
}[dxd\theta]$ and $\left[  d\left(  dx\,\right)  d\left(  d\theta\right)
\right]  \rightarrow\lambda\left[  d\left(  dx\,\right)  d\left(
d\theta\right)  \right]  . $ This can be extended to general coordinate
transformations, and the outcome is that $[d^{n}x\,d^{m}\theta d^{n}\left(
dx\,\right)  d^{m}\left(  d\theta\right)  ]$ is an \textbf{invariant} measure.

It is clear now that \textbf{we cannot integrate a generic }$\omega\left(
x,\theta,dx,d\theta\right)  .$ Suppose that the Riemann-Lebesgue integrability
conditions are satisfied with respect to the $x$ variables; the integrals over
$dx$ and $\theta$ (being Berezin integrals) pose no further problem but, if
$\omega\left(  x,\theta,dx,d\theta\right)  $ has a polynomial dependence in
the (bosonic) variables $d\theta$, the integral diverges unless $\omega\left(
x,\theta,dx,d\theta\right)  $ depends on the $d\theta$ only through the
product of \textbf{all} the ``distributions" $\delta\left(  d\theta^{\alpha
}\right) $.\footnote{We could (as explained above) also admit a more general
$d\theta$ dependence i.e. in the form of a test function in the $d\theta
^{\alpha}$ multiplied by the product of all distributions $\delta\left(
d\theta^{\alpha}\right)  ,$ but this generalization is not needed here.} This
solves the problem of the divergences for all the $d\theta^{\alpha}$ variables
because
\begin{equation}
\int\delta\left(  d\theta^{\alpha}\right)  [d\left(  d\theta^{\alpha}\right)
]=1
\end{equation}
Summing up we can integrate only integral forms $\omega$, the integral
selecting only forms contained in $\omega$ with top degree in bosonic
variables and top picture number, namely the so-called \textbf{integral top
forms}.

In order to shorten notations, when the ``variables of integration" are
evident, we will omit in the integrals all the ``integration measures symbols"
such as $[d^{n}x\,d^{m}\theta d^{n}\left(  dx\,\right)  d^{m}\left(
d\theta\right)  ]$ or $[d^{n}x\,d^{m}\theta]$.

In the case of curved supermanifolds, by expressing the $1$-forms $dx^{i}$ and
$d\theta^{\a}$ in terms of the supervielbeins $E_{M}^{A}\equiv(e^{i}_{M},
e^{\alpha}_{M})$ (where $A$ runs over the flat indices $i$ and $\alpha$, and
$M$ runs over the curved indices), we have
\begin{equation}
\int_{\Omega^{\bullet}(\mathcal{M})}\omega(x,\theta,e^{i},e^{\a}%
)=\int_{\mathcal{M}}E\,\epsilon^{i_{1}\dots i_{n}}\epsilon^{\alpha_{1}%
\dots\alpha_{m}}\,\omega_{[i_{1}\dots i_{n}][\alpha_{1}\dots\alpha_{m}%
]}(x,\theta)\label{inM}%
\end{equation}
where $E=\mathrm{sdet}(E_{M}^{A})$ is the superdeterminant (the Berezinian) of
the supermatix $E_{M}^{A}(x,\theta)$. As usual this definition is invariant
under (orientation preserving super) diffeomorphisms.

%%%%%%%%%%%%%%%%%%%%%%%%%%%%%%%%%%%%%%%%%%%%%%%%%%%%%%%%%%%

\section{Poincar\'{e} duals and Variational Principles on Submanifolds}

As discussed in the introduction, we consider a submanifold $\mathcal{S}$ of a
bigger space $\mathcal{M}$ -- that could be also a supermanifold -- and we
give a recipe to construct an action $I$ on that submanifold. The next step is
to derive the equations of motion from a variational principle varying both
the Lagrangian $L$ and the embedding of the submanifold into $\mathcal{M}$.
This can be achieved by extending the integral of the Lagrangian $L$ to an
integral over the entire bigger space $\mathcal{M}$. For that we need the
notion of the Poincar\'e dual of the submanifold $\mathcal{S}$ into
$\mathcal{M}$. The result is an extended Lagrangian, depending dynamically on
the fields and on the embedding functions, integrated over a fixed manifold
$\mathcal{M}$.

\subsection{Poincar\'e Duals}

We start with a submanifold $\mathcal{S}$ of dimension $s$ of a differentiable
manifold $\mathcal{M}$ of dimension $n.$ We take an embedding $i:$
%L%
\[
i:\mathcal{S}\rightarrow\mathcal{M}%
\]
and a compact support form $L\in\Omega^{s}(\mathcal{M}).$ The
\textbf{Poincar\'{e} dual} of $\mathcal{S}$ is a \textit{closed} form
$\eta_{\mathcal{S}}\in\Omega^{n-s}(\mathcal{M})$ such that $\forall L$:%
\begin{equation}
I[L,\mathcal{S}]=\int_{\mathcal{S}}i^{\ast}L=\int_{\mathcal{M}}L\wedge
\eta_{\mathcal{S}}\label{catA}%
\end{equation}
where $i^{\ast}$ is the pull-back of forms. We are not interested here in a
rigorous mathematical treatment (see \cite{bott-tu}) and we take a heuristic
approach well-adapted for the generalization to the supermanifold case. In the
symbol $I[L,\mathcal{S}]$, we have recalled the dependence upon the embedding
of $\mathcal{S}$ into $\mathcal{M}$.

If we suppose that the submanifold $\mathcal{S}$ is described locally by the
vanishing of $n-s$ coordinates $t^{1},\dots,t^{n-s}$, its Poincar\'{e} dual
can also be described as a \textit{singular closed localization form} (the
correct mathematics is the de Rham current theory \cite{griffith-harris}):%
\begin{equation}
\eta_{_{\mathcal{S}}}=\delta(t^{1})...\delta(t^{n-s})dt^{1}{}\wedge...{}\wedge
dt^{n-s}\label{catB}%
\end{equation}
This distribution-valued form is clearly closed (from the properties of the
delta distributions $d\,\delta(t)=\delta^{\prime}(t)dt$ and from $dt^{i}\wedge
dt^{i}=0$). This form belongs to $\Omega^{n-s}(\mathcal{M)}$ and is
constructed in such a way that it projects on the submanifold $t^{1}%
=\dots=t^{n-s}=0$ and orthogonally to $dt^{1}{}\wedge\dots{}\wedge dt^{n-s}$.
Thus, by multiplying a given form $L \in\Omega^{s}(\mathcal{M)}$ by $\eta_{S}%
$, the former is restricted to those components which are not proportional to
the differentials $dt^{i}$.

Observing that the Dirac $\delta$-function of an odd variable ($dt$ is odd if
$t$ is even) coincides with the variable itself (as can be seen using Berezin
integration), we rewrite $\eta_{\mathcal{S}}$ as a form that will turn out to
be very useful for generalization (omitting wedge symbols):
\begin{equation}
\eta_{_{\mathcal{S}}}=\delta(t^{1})...\delta(t^{n-s})\delta(dt^{1}){}%
...{}\delta(dt^{n-s})\label{catC}%
\end{equation}
which heuristically corresponds to the localisation to $t^{1}=\dots=t^{n-s}=0
$ and $dt^{1}=\dots=dt^{n-s}=0$. Note that if a submanifold $\mathcal{S}$ is
described by the vanishing of $n-s$ functions $f^{1}(t)=\dots=f^{n-s}(t)=0$
the corresponding Poincar\'{e} dual $\eta_{_{\mathcal{S}}}$ is:
\[
\eta_{_{\mathcal{S}}}=\delta(f^{1})...\delta(f^{n-s})\delta(df^{1}){}%
...\delta(df^{n-s})
\]
This form, when written completely in terms of the $t^{i}$ coordinates,
contains also the \textit{derivatives of the }$\delta$'s because of the
expansion of $\delta(f)$ and $\delta(df)$ in terms of $t^{i}$.

If we change (in the same homology class) the submanifold $\mathcal{S}$ to
$\mathcal{S}^{\prime}$ the corresponding Poincar\'{e} duals $\eta
_{_{\mathcal{S}}}$ and $\eta_{_{\mathcal{S}^{^{\prime}}}}$ are known to differ
by an \textit{exact} form:%
\[
\eta_{_{\mathcal{S}}}-\eta_{_{\mathcal{S}^{^{\prime}}}}=d\gamma
\]
This can be easily proved by recalling that the Poincar\'{e} duals are closed
$d\eta_{S}=0$ and any variation (denoted by $\Delta$) of $\eta_{\mathcal{S}}$
is exact:
\begin{equation}
\Delta\eta_{_{\mathcal{S}}}=d\Big(\Delta f\delta(f)\Big)\label{catD}%
\end{equation}
Given the explicit expression of $\eta_{_{\mathcal{S}}}$, it is easy to check
eq. (\ref{catD}) by expanding both members (assuming that $\Delta$ follows the
Leibniz rule) and using the distributional laws of $\delta$'s.

Using this property we can show that, if $dL=0$ (in $\mathcal{M}$ since
$d_{\mathcal{S}}\left(  i^{\ast}L\right)  =0$ trivially in $\mathcal{S}$),
then the action does not depend on the embedding of  the submanifold. Indeed
varying the embedding amounts to vary the Poincar\'e dual, so that the
variation of the integral reads
\begin{equation}
\label{stokke}\Delta I[L, \mathcal{S}] = I[L, \Delta\mathcal{S}] =
\int_{\mathcal{M}} L \wedge\Delta\eta_{\mathcal{S}} = \int_{\mathcal{M}}
L\wedge d \xi_{\mathcal{S}} = (-)^{s} \int_{\mathcal{M}} d L \wedge
\xi_{\mathcal{S}}%
\end{equation}
where $\Delta\eta_{\mathcal{S}} = d \xi_{\mathcal{S}}$.

The same arguments apply in the case of supermanifolds. Consider a submanifold
$\mathcal{S}$ of dimension $s|q$ of a supermanifold $\mathcal{M}$ of dimension
$n|m.$ We take an embedding $i:$%
\[
i:\mathcal{S}\rightarrow\mathcal{M}%
\]
and an integral form $L \in\Omega^{s|q}(\mathcal{M})$ (integrable in the sense
of superintegration when pulled back on $\mathcal{S}$). The
\textbf{Poincar\'{e} dual} of $\mathcal{S}$ is a $d$-\textit{closed} form
$\eta_{\mathcal{S}}\in\Omega^{n-s|m-q}(\mathcal{M})$ such that:%
\[
\int_{\mathcal{S}}i^{\ast}L =\int_{\mathcal{M}}L \wedge\eta_{\mathcal{S}}%
\]
Again we can write:%
\[
\eta_{_{S}}=\delta(f^{1})...\delta(f^{...})\delta(df^{1})...\delta(df^{...})
\]
where the $f$'s are the functions defining (at least locally) the submanifold
$\mathcal{S}$. Here some of them are even functions and some of them are odd
functions, accordingly the Poincar\'{e} dual is a closed integral form that,
written in the coordinates $(x,\theta)$, contains delta forms and their derivatives.

Again it is easy to check that any variation of $\eta_{S}$ is $d$-exact:
\begin{equation}
\Delta\eta_{S}=d\Big((\Delta f)f\delta^{^{\prime}}(df)\Big)\label{catD-1}%
\end{equation}
Note that the two formulae (\ref{catD}) and (\ref{catD-1}) for the variation
of $\eta_{S}$ can be combined in a formula that holds true in both cases:
\begin{equation}
\Delta\eta_{S}=d\Big(\Delta f\delta(f)\delta^{^{\prime}}%
(df)\Big)\label{catD-1-1}%
\end{equation}
Indeed, one has $\delta^{^{\prime}}(df)=1$ or $\delta(f)=f$ when $f$ is
respectively bosonic or fermionic.

Before considering some examples, we have to spend a few words on the general
form of the Poincar\'e dual in the case of supermanifolds:
\begin{equation}
\label{thomclassB}{\ \eta^{(n-s|m)}_{\mathcal{S}} = \sum_{l=0}^{d}
\eta_{_{[i_{1} \dots i_{n-s +l}]}}(x, \theta) dx^{i_{1}} \dots dx^{i_{n-s +
l}} \partial^{l} \delta^{m}(d\theta) }%
\end{equation}
where we have added $l$-derivatives $\partial^{l}$ on the Dirac delta
functions (for the moment we have not specified how these derivatives are
distributed on $\delta^{m}(d\theta)$, but we have to admit all possible
combinations and, for each of them, we have new coefficients $\eta_{_{[i_{1}
\dots i_{n-s +l}]}}(x, \theta)$). Acting with derivatives on Dirac delta's we
decrease the form number which must be compensated by adding more 1-forms $dx$
up to the maximum $n$ (this implies also that the maximum number of
derivatives is $s$). In principle, we could have also added $d\theta$, but
these can be removed by integration by parts. Notice that the simple
Poincar\'e dual given in (\ref{catC}) is included in the general expression
(\ref{thomclassB}). If we consider again the integral  with ${L}$, by
integration-by-parts and by using the property $d\theta\delta^{\prime}%
(d\theta) = - \delta(d\theta)$, we can finally take into account all possible
directions (namely also the $d\theta$ directions). We would like also to
underline that the different coefficients $\eta_{_{[i_{1} \dots i_{n-s +l}]}%
}(x, \theta)$ parametrize all possible embeddings of the submanifold
$\mathcal{S}$ into the supermanifold $\mathcal{M}^{(n|m)}$. In particular they
parametrize how the coordinates of the submanifold are written in terms of
those of the complete supermanifold and this amounts to the choice of
derivatives of Dirac delta's.

Let us consider for example $\mathbb{R}^{(0|1)}$ as a submanifold of
$\mathbb{R}^{(0|2)}$, which has two coordinates $\theta^{1}$ and $\theta^{2}$.
The form $L=\theta^{1}\delta(d\theta^{2})\in\Omega^{(0|1)}$ can be integrated
over the submanifold $\mathbb{R}^{(0|1)}$ since it is a $0-$form with
$1-$picture. The embedding of $\mathbb{R}^{(0|1)}$ is chosen by setting
$a\,\theta^{1}+b\,\theta^{2}=0$ with $a,b\in\mathbb{R}$. We can compute the
integral in two ways: the first is by using $\theta^{1}=-\frac{b}{a}\theta
^{2}$ and by re-expressing $L$ in terms of the coordinate $\theta^{2}$ only.
Thus
\[
L=-\frac{b}{a}\theta^{2}\delta(d\theta^{2})
\]
and the integral gives:
\[
I[L,(a,b)]=\int_{\mathbb{R}^{(0|1)}}L=-\frac{b}{a}%
\]
The second way is as follows. The Poincar\'{e} dual of $\mathbb{R}^{(0|1)}$
into $\mathbb{R}^{(0|2)}$ is
\[
\eta_{S}=\delta(a\,\theta^{1}+b\,\theta^{2})\delta(a\,d\theta^{1}%
+b\,d\theta^{2})\,.
\]
The first delta function can be rewritten as $a\,\theta^{1}+b\,\theta^{2}$
because of the anticommutativity of $\theta$'s. Multiplying $\eta_{S}$ by $L$
we obtain:
\[
L\wedge\eta_{S}=\theta^{1}\delta(d\theta^{2})\delta(a\,\theta^{1}%
+b\,\theta^{2})\delta(a\,d\theta^{1}+b\,d\theta^{2})=
\]%
\[
=\theta^{1}(a\,\theta^{1}+b\,\theta^{2})\delta(d\theta^{2})\delta
(a\,d\theta^{1}+b\,d\theta^{2})=
\]%
\[
=b\theta^{1}\theta^{2}\delta(d\theta^{2})\delta(a\,d\theta^{1})=-\frac{b}%
{a}\theta^{1}\theta^{2}\,\delta(d\theta^{1})\delta(d\theta^{2})
\]
Thus $\int_{\mathbb{R}^{(0|2)}}L\wedge\eta_{S}=-b/a$ which coincides with the
computation above. The integral depends upon the embedding parameters $(a,b)$
(and is not defined for $a=0$). Repeating the same computation with a closed
form (for example $\theta^{1}\delta(d\theta^{1})$), it is easy to see that the
integral equals $1$ and does not depend on the embedding parameters as expected.

\subsection{Variational Principle}

The action $I[L, \mathcal{S}]$ is a functional of $L$ and $\mathcal{S}$, and
therefore varying it means varying both $L$ and $\mathcal{S}$. The latter
corresponds to varying $\eta_{S}$. The variational principle leads to
\begin{equation}
\label{varpri}{\ \Delta I[L, \mathcal{S}] = \Delta\int_{\mathcal{S}} i^{*} L=
\int_{\mathcal{M}} ( \Delta L\wedge\eta_{S} + L\wedge\Delta\eta_{S}) } =0\,.
\end{equation}
The variation has two terms. The first one contains the variation of the
Lagrangian $L$ over the entire space and the second one the variation of the
embedding. However, in the second term we use the exactness of the variation
of $\eta_{S}$ ($\Delta\eta_{S} = d \xi_{S}$) and by integration by parts we
can rewrite the variation of the action as
\begin{equation}
\label{varpriB}\Delta I[L, \mathcal{S}] = \int_{\mathcal{M}} ( \Delta
L\wedge\eta_{S} + (-)^{s} dL\wedge\xi_{S})
\end{equation}
where $s$ is the degree of the form $L$. The expression $\xi_{S}$ is arbitrary
since it corresponds to an arbitrary variation of $\mathcal{S}$, and therefore
both terms of the integral must vanish separately leading to the equations of
motion
\begin{equation}
\label{varpriC}\Delta L= 0\,, ~~~~~~ d L= 0 \,.
\end{equation}
Since the variation of $L$ under $\Delta$ is an arbitrary variation, the first
equation implies the second one and therefore, on the equations of motion
$\Delta L =0$ (only), the integral $I[L, \mathcal{S}]$ is independent of
$\mathcal{S}$. This is somehow rather obvious, but it is interesting to notice
that in many cases $dL= 0$ holds only on a subset of the equations of motion,
and in some cases it holds completely off-shell.

As an example we consider $3d-$euclidean gravity on a $3d-$submanifold $S$
(for example a $3d-$topological sphere) embedded into $\mathbb{R}^{4}$. The
Poincar\'{e} dual is given by $\eta_{S}=\delta(f)df$ where $S=\left\{
f^{-1}\left(  0\right)  \right\}  $. The action is given by
\begin{equation}
I[\omega,V,f]=\int_{S}i^{\ast}\epsilon_{abc}R^{ab}(\omega)\wedge V^{c}%
=\int_{\mathbb{R}^{4}}\epsilon_{abc}R^{ab}(\omega)\wedge V^{c}\wedge
\delta(f)df\,,\label{exavarA}%
\end{equation}
where $\omega$ is the spin connection, $R^{ab}=d\omega^{ab}-\omega_{~c}%
^{a}\wedge\omega^{cb}$, $V^{a}$ is the \textit{dreibein} and $f$ is the
embedding function. The equations of motion are given by
\begin{align}
& R^{ab}\delta(f)\wedge df=0\,,~~~~~T^{a}\delta(f)\wedge
df=0\,,~~~~\nonumber\\
&  d\left(  \epsilon_{abc} R^{ab} \wedge V^{c} \right)  = \epsilon_{abc}%
R^{ab}(\omega)\wedge T^{c} = 0\,.\label{exavarB}%
\end{align}
where the torsion is defined as $T^{a} = d V^{a} + \omega^{a}_{~b} \wedge
V^{b}$. Notice that the equation on the first line are valid for any $f$, and
this implies that $R^{ab}=0,T^{a}=0$ on the entire space $\mathbb{R}^{4}$. The
last equation is a consequence of the first two equations together with the
Bianchi identity $D R^{ab} =0$ (where $D$ is the Lorenz covariant derivative),
but we observe that only one is sufficient to guarantee the vanishing of the
last equation. Namely, for a torsionless connection $\omega$, $dL=0$ off-shell
and $\omega$ can be expressed in terms of $V^{a}$ (second order formalism).

We consider now $4d$-Einstein gravity and we would like to embed the
Einstein-Hilbert Lagrangian (defined on 4-dimensional space $\mathcal{S}$) in
a bigger 10-dimensional space $\mathcal{M}$ viewed as the group manifold
associated to Poincar\'e symmetry generated by the translations and by Lorentz
transformations. The coordinates of $\mathcal{M}$ are the usual $x^{a}$ and
the ``Lorentz coordinates'' $y^{ab}$. The exterior bundle $\Omega^{\bullet
}(\mathcal{M)}$ is parametrised by the vielbeins $V^{a}, \omega^{ab}$ (they
are interpreted as the usual $4d$-vielbein and the spin connection). The
curvatures $T^{a}$ (associated to the translations) and $R^{ab}$ (associated
to the Lorentz transformations) can be decomposed along the complete basis
\begin{align}
& T^{a} = T^{a}_{~b} V^{b} + T^{a}_{~bc} \omega^{bc}\,,\nonumber\\
& R^{ab}=R^{ab}_{~~cd}V^{c} \wedge V^{d} + R^{ab}_{~~c,de}V^{c} \wedge
\omega^{de}+ R^{ab}_{~~cd,ef}\omega^{cd} \wedge\omega^{ef}%
\end{align}
We denote by \textit{inner} components the coefficients along $V^{a}\wedge
V^{b}$ and \textit{outer} the remaining ones.

The EH action is written as
\begin{equation}
I_{EH} [\omega,V]=\int_{\mathcal{S}} i^{*} \left(  R^{ab} \wedge V^{c} \wedge
V^{d} \epsilon_{abcd} \right)  = \int_{\mathcal{M}} R^{ab} \wedge V^{c} \wedge
V^{d} \epsilon_{abcd} \wedge\eta_{\mathcal{S}}%
\end{equation}
where $\eta_{\mathcal{S}}$ is the Poincar\'e dual of $\mathcal{S}$ in
$\mathcal{M}$. Under the conditions discussed above the equations of motion on
the big space $\mathcal{M}$ are
\begin{equation}
\epsilon_{abcd} T^{c} \wedge V^{d} \wedge\eta_{\mathcal{S}}=0,~~~
\epsilon_{abcd} R^{ab} \wedge V^{d} \wedge\eta_{\mathcal{S}} = 0
\end{equation}
The field equations are 3-form equations on $\mathcal{M}$. Their content can
be extracted by projecting on a complete basis of 3-forms in $\mathcal{M}$.
The first equation is then found to imply $T^{a}=0$ (i.e. the torsion vanishes
as a 2-form on $\mathcal{M}$), and the second leads to the vanishing of the
outer components of the Lorentz curvature $R^{ab}$ and to the Einstein
equations for the inner components $R^{ab}_{~~cd}$:
\begin{equation}
R^{ac}_{~~bc} - \frac{1}{2} \delta^{a}_{~b} R^{cd}_{~~cd} =0
\end{equation}
It is easy to check that the field equations imply
\[
d ( R^{ab} \wedge V^{c} \wedge V^{d} \epsilon_{abcd} )=0
\]
by using $d L=D L$ (the covariant exterior derivative for any Lorentz
invariant quantity $L$), the Bianchi identity $D R^{ab}=0$ and the field
equation $T^{a}=0$.

\subsection{Invariances of the Action}

By construction, integrals of top forms are invariant under (infinitesimal)
diffeomorphisms (hereafter simply called diffeomorphisms). Indeed  the action
of a Lie derivative $\mathcal{L}_{\epsilon}$ along a tangent vector $\epsilon$
on a top form $\Omega$ is a total derivative $d (\iota_{\epsilon}\Omega)$. We
consider the Lagrangian $L(\mu)$ as a function of the fields $\mu$ which are
$p$-forms, their wedge products and their exterior derivative. Thus one knows
a priori that the action
\begin{equation}
I[L(\mu), \mathcal{S}] = \int_{\mathcal{M}} L(\mu) \wedge\eta_{\mathcal{S}%
}\label{actionG}%
\end{equation}
is invariant under diffeomorphisms in $\mathcal{M}$, $L\wedge\eta
_{\mathcal{S}}$ being a top form.  The variation of this action under
diffeomorphisms can be written again as a sum of two pieces:
\begin{equation}
\delta_{\epsilon}I = 0 = \int_{\mathcal{M}} \mathcal{L}_{\epsilon}L \wedge
\eta_{\mathcal{S}} + \int_{\mathcal{M}} L \wedge\mathcal{L}_{\epsilon}%
\eta_{\mathcal{S}}\label{generaldiff}%
\end{equation}
Consider first $\epsilon$ to be a tangent vector lying along $\mathcal{S}$, so
that we are dealing with diffeomorphisms  in the $\mathcal{S}$ submanifold (we
call them for short $x$-diff's). The $x$-diff's do not change the embedding of
$\mathcal{S}$ inside $\mathcal{M}$,  and therefore leave unchanged the
Poincar\'e dual $\eta_{\mathcal{S}}$. Thus for $x$-diff's
\begin{equation}
\delta_{\epsilon}I = 0 = \int_{\mathcal{M}} \mathcal{L}_{\epsilon}L\wedge
\eta_{\mathcal{S}}\label{xdiff}%
\end{equation}
and we find that varying \textit{only the Lagrangian $L$} under $x$-diff's
leaves the action (\ref{actionG}) invariant. Since the fields $\mu$ appear
only in the Lagrangian, \textit{the action $I$ is invariant under $x$-diff's
variations of the fields $\mu$}.  In terms of the embedding $i$, the variation
of $I$ under $x$-diff's given in eq. (\ref{xdiff}) can be written as
\begin{equation}
\delta_{\epsilon}\int_{\mathcal{S}} i^{*} L = 0 = \int_{\mathcal{S}} i^{*} (
\mathcal{L}_{\epsilon}L )
\end{equation}
The l.h.s. corresponds to a variation of the fields in $i^{*}(L)$, the r.h.s.
corresponds to a diffeomorphism variation of L.

The situation is different when the tangent vector $\epsilon$ lies outside the
tangent space of  $\mathcal{S}$ (we call the corresponding diffeomorphisms
$y$-diff's). In this case, the embedding  of $\mathcal{S}$ inside
$\mathcal{M}$ changes under the action of the Lie derivative $\mathcal{L}%
_{\epsilon}$, and the second term of eq.  (\ref{generaldiff}) is present.
Recalling that $d \eta_{\mathcal{S}}=0$, this term reduces, after integration
by parts, to  $(-)^{s} \int_{\mathcal{M}} dL \wedge\iota_{\epsilon}\eta_{S}$.
Thus varying only $L$ under $y$-diff's leaves the action (\ref{actionG})
invariant if $dL=0$.  We conclude that $y$-diff's applied to the fields $\mu$
are invariances of the action $I$ if $dL=0$.

Actually the condition for $y$-diff's on $\mu$ to be invariances of $I$ is
weaker: indeed it is sufficient to have $\iota_{\epsilon}dL=0$.  This can be
checked directly by varying $L$ in the action under $y$-diff.s:
\begin{equation}
\label{lastpippo}\int_{\mathcal{M}} \mathcal{L}_{\epsilon}L \wedge
\eta_{\mathcal{S}} = \int_{\mathcal{M}} [(\iota_{\epsilon}d L) \wedge
\eta_{\mathcal{S}} + d ( \iota_{\epsilon}L) \wedge\eta_{\mathcal{S}}]
\end{equation}
Integrating by parts the second term and recalling that $d\eta_{\mathcal{S}%
}=0$ \textit{proves that the action $I$  is invariant under $y$-diff's applied
to the fields $\mu$} when $\iota_{\epsilon}dL=0$.

%We have seen that the condition $dL =0$ is sufficient for the action $I[\mu, {\cal S}]$ to be independent on the
%embedding. However, in same cases, this condition can be weakened.
%We want a dynamical theory, living on ${\cal S}$, for the fields $\mu$ contained in the Lagrangian.
%However we have seen that the action $I$ also depends on the embedding functions describing how ${\cal S}$ is contained in $\Gtilde$. Can we remove this dependence?
%It turns out that

The last equation (\ref{lastpippo}) can also be used to study the dependence
of the action upon the embedding functions. We know  that $\int_{\mathcal{M}}
L \wedge\mathcal{L}_{\epsilon}\eta_{\mathcal{S}} = - \int_{\mathcal{M}}
\mathcal{L}_{\epsilon}L \wedge\eta_{\mathcal{S}} $ from eq. (\ref{generaldiff}%
). Thus any variation of the embedding (generated by $\mathcal{L}_{\epsilon}$,
with an arbitrary $\epsilon$  outside $\mathcal{S}$) can be compensated by a
$y$-diff's on $L$. On the other hand we have seen that $y$-diff's on $L$  do
not change the action when $\iota_{\epsilon}dL=0$ with $\epsilon$ in the
$y$-directions,  and therefore this is also the condition for $I$ to be
independent on the particular embedding of $\mathcal{S}$.

%Note that the submanifold of $\Gtilde$ we are considering is not the most general one, but is a smooth
%deformation of a coset space $G/H$ where $H$ is given (in most cases it contains a Lorentz subgroup of $G$).
%This is the reason we only need  $\iota_t dL=0$ rather than $dL=0$ (see Section 3) to have independence on the submanifold embedding.

Let us come back to our example, pure gravity in the group manifold approach,
where the ``big space" $\mathcal{M}$  is (a smooth deformation of) the
Poincar\'e group manifold, and the ``small space" $\mathcal{S}$ is the usual
Minkowski spacetime. Usual $x$-diff's on the fields $V^{a}$ and $\omega^{ab}$
leave the action invariant, while $y$-diff's, i.e. diffeomorphisms along the
Lorentz directions of $\mathcal{M}$, are invariances when applied to $V^{a}$
and $\omega^{ab}$ if $\iota_{t} dL=0$ ($t = t^{ab} \partial_{y^{ab}}$ being
the tangent vectors in the Lorentz directions, dual to the spin connection
$\omega^{ab}$). Let us check whether this condition holds. Replacing again the
exterior derivative $d$ with the covariant exterior derivative $D$,  and using
the Bianchi identity $D R^{ab}=0$ and definition of the torsion, we find the
condition:
\begin{equation}
\iota_{t }d L=\iota_{t} (R^{ab} \wedge T^{c} \wedge V^{d})~\epsilon_{abcd} =0
\end{equation}
Using now the Leibniz rule for the contraction, and $\iota_{t }(V^{a})=0$,
leads to the condition that  all outer components of $R^{ab}$ and $T^{a}$ must
vanish. These conditions are \textit{part} of the field  equations previously
derived. In particular they do not involve the ``inner" field equations, i.e.
the  Einstein equations. On this ``partial shell" the action is invariant
under $y$-diff's (``Lorentz diffeomorphisms")  applied to the fields.

The vanishing of outer components of the curvature is also called
\textit{horizontality} of the curvature.

When horizontality of $R^{A}$ in the $y$-directions holds, the dependence of
fields $\mu^{A} (x,y)$ on $y$ is completely determined by their value $\mu
^{A}(x,0)$ on the embedded hypersurface $\mathcal{S}$. Indeed in this case an
infinitesimal $y$-diffeomorphism on $\mu^{B} (x,0)$ can be written as
\begin{equation}
\mu^{B}(x,\delta y) = \mu^{B}(x,0) + d \delta y^{B} + C^{B}_{~CD} \mu^{C}(x,0)
\delta y^{D}%
\end{equation}
and shows that $\mu^{B}(x,\delta y)$ is determined by the value of the field
$\mu$ at $y=0$. This equation can be integrated to reconstruct the
$y$-dependence of $\mu^{B}(x, y)$ (at least in a sufficiently small connected
neighborhood of $y=0$).

A milder form of horizontality occurs when the outer components of the
curvature $R^{A}$ do not vanish, but are proportional to linear combinations
of inner components of $R^{A}$. The curvature is then said to be
\textit{rheonomic}. In this case a diffeomorphism in the outer directions
involves only the values of $\mu^{A}(x,0)$, and of its $x$-space derivatives
$\frac{\partial}{\partial x^{\mu}} \mu^{A}(x,0)$, contained in the inner
components of $R^{A}$. Again the value of $\mu^{A}(x,0)$ on the hypersurface
$\mathcal{S}$ determines the $y$-dependence of $\mu^{A}(x,y)$ on the manifold
$\mathcal{M}$. This situation is very common in supergravity theories, where
some of the outer directions are fermionic, and diffeomorphisms in these
directions are interpreted as supersymmetry transformations.

\subsection{Field transformation rules}

Let us have a closer look at the variation of the fields $\mu$ under
(infinitesimal) diffeomorphisms.  The transformation rule is given by the
action of the Lie derivative on $\mu$:
\begin{equation}
\delta_{\epsilon}\mu= \mathcal{L}_{\epsilon}\mu= d \iota_{\epsilon}\mu+
\iota_{\epsilon}d \mu\label{muLie}%
\end{equation}
When $\mu^{A}$ is the vielbein of a (deformed) group manifold $\mathcal{M}$
(the index $A$ running on the Lie algebra of $G$),  the variation formula
(\ref{muLie}) takes the suggestive form:
\begin{equation}
\delta_{\epsilon} \mu^{B} = d \epsilon^{B} + C^{B}_{~CD} \mu^{C} \epsilon^{D}
+ \iota_{\epsilon} R^{B} \equiv(\nabla\epsilon)^{B} + \iota_{\epsilon}
R^{B}\label{mutransf}%
\end{equation}
where $C^{B}_{~CD}$ are the $G$-structure constants, $\epsilon= \epsilon^{A}
t_{A}$  is a generic tangent vector expanded on the tangent basis $t_{A}$ dual
to the  cotangent (vielbein) basis $\mu^{B}$, and $\nabla$ is the
$G$-covariant exterior derivative.  To prove this one just uses the definition
of the group curvatures:
\begin{equation}
R^{A} = d \mu^{A} + \frac{1}{2} C^{A}_{~BC} \mu^{B} \wedge\mu^{C}\label{Rdef}%
\end{equation}
that allow to re-express $d\mu^{A}$ in terms of $R^{A}$ and bilinears of vielbeins.

When the group curvatures $R^{A}$ are horizontal in the directions of some
subgroup $H$ of $G$, the diffeomorphisms along the $H$-directions become
\textit{gauge transformations}, as one sees immediately from the
diffeomorphism variation formula (\ref{mutransf}): indeed in this case the
contracted curvature term vanishes, and the variation amounts to the covariant
derivative of the parameter $\epsilon$. Thus the group-geometric approach
provides a unified picture of the symmetries (gauge or diffeomorphisms): they
all originate from diffeomorphism invariance in $\mathcal{M}$.

In our example of pure gravity where $\mathcal{M}$ is a deformed Poincar\'e
manifold, the $y$-diff's transformation rules (\ref{mutransf}) are obtained by
choosing the tangent vector $\epsilon$ in the Lorentz directions, $\epsilon=
\epsilon^{ab} t_{ab}$, and by using the horizontality of $T^{a}$ and $R^{ab}$
in the Lorentz directions. One finds
\begin{equation}
\delta_{\epsilon}V^{a} = \epsilon^{a}_{~b} V^{b},~~~\delta_{\epsilon}%
\omega^{ab} = D \epsilon^{ab} \equiv d \epsilon^{ab} - \omega^{a}_{~c}
\epsilon^{cb}+ \omega^{b}_{~c} \epsilon^{ca}%
\end{equation}
reproducing the Lorentz gauge variations of the vielbein and the spin
connection. The infinitesimal parameter of the diffeomorphism transformation
$\epsilon^{ab}$ in the Lorentz coordinates is then re-interpreted as the local
Lorentz gauge parameter. The Einstein-Hilbert action on $\mathcal{S}$ is
invariant under these transformations.

\subsection{Supersymmetry}

In the group-geometric approach to supergravity theories, the ``big" manifold
$\mathcal{M}$ is a supergroup manifold, and there are fermionic vielbeins
$\psi$ (the gravitini) dual to the fermionic tangent vectors in $\mathcal{M}$.
The diffeomorphisms in the fermionic directions are a particular instance of
the general rule (\ref{mutransf}). When rheonomy holds, the fermionic
diffeomorphisms are seen as (local) supersymmetry variations of the fields. To
illustrate this mechanism, we consider the example of $D=4$ simple
supergravity, for which $G$ is the superPoincar\'e group. The fields $\mu^{A}$
are in this case the vielbein $V^{a}$, the gravitino (a Majorana 1-form
fermion) $\psi$, and the spin connection $\omega^{ab}$ corresponding
respectively to the translations, supersymmetries and Lorentz rotations of the
superPoincar\'e Lie algebra. The general curvature definition (\ref{Rdef})
becomes, using the structure constants of the Lie superalgebra:
\begin{equation}
T^{a}= dV^{a} - \omega^{ac} V^{c} - \frac{i }{2} \bar\psi\gamma^{a} \psi,~~
\rho= d \psi- \frac{1}{ 4} \omega^{ab} \gamma_{ab} \psi,~~ R^{ab}=d
\omega^{ab} - \omega^{ac} \omega^{cb}%
\end{equation}
defining respectively the supertorsion, the gravitino field strength and the
Lorentz curvature. All forms live on $\mathcal{M}$ = (deformed)
super-Poincar\'e group manifold. The action is a $4$-form integrated on a
$\mathcal{S}$ (diffeomorphic to Minkowski spacetime) submanifold of
$\mathcal{M}$:
\begin{equation}
I[V,\omega,\psi]= \int_{\mathcal{S}} R^{ab} V^{c} V^{d} \epsilon_{abcd} + 4
\bar\psi\gamma_{5} \gamma_{a} \rho V^{a}%
\end{equation}
The field equations, when projected on all the $\mathcal{M}$ directions, give
the following conditions on the curvatures:
\begin{align}
&   R^{ab} = R^{ab}_{~~cd} V^{c} \wedge V^{d} -(\epsilon_{abcd} \bar\rho_{cd}
\gamma_{5} \gamma_{e} + \delta^{[a}_{e} \epsilon^{b]cdf} \bar\rho_{df}
\gamma_{5} \gamma_{c}) \wedge\psi\wedge V^{e}\label{cond1}\\
&   T^{a} =0\label{cond2}\\
&   \rho= \rho_{ab}V^{a}\wedge V^{b}\label{cond3}%
\end{align}
where the spacetime (inner) components $R^{ab}_{~~cd}$, $\rho_{ab}$ satisfy
the propagation equations
\begin{equation}
R^{ac}_{~~bc} - \frac{1}{ 2} \delta^{a}_{~b} R^{cd}_{~~cd} =0,~~\gamma^{abc}
\rho_{bc}=0\label{prop}%
\end{equation}
respectively the Einstein and the gravitino field equations. Eq.s
(\ref{cond1})-(\ref{cond3}), an output of the equations of motion, are
rheonomic conditions. Indeed the only nonvanishing outer components (those of
$R^{ab}$) are given in terms of the inner components $\rho_{ab}$.

The symmetries of the theory are encoded in the general diffeomorphism formula
(\ref{mutransf}), and are given by ordinary $\mathcal{S}$ diffeomorphisms,
local Lorentz rotations (diff.s in the Lorentz directions) and local
supersymmetry transformations (diff.s in the fermionic directions). The latter
read:
\begin{equation}
\delta_{\epsilon}V^{a} = i \bar\epsilon\gamma^{a} \psi,~~~\delta_{\epsilon
}\omega^{ab} = 2 \bar\theta^{ab}_{~~c} \epsilon V^{c},~~~\delta_{\epsilon}%
\psi= d\epsilon-{\frac{1 }{4}} \omega^{ab} \gamma_{ab} \epsilon
\end{equation}
where $\bar\theta^{ab}_{~~c}$ are the $\psi V^{c}$ components of $R^{ab}$
given in (\ref{cond1}).

At this juncture, one may wonder whether the action is invariant under
supersymmetry transformations: as discussed in a previous subsection, this
will be the case if the contraction of $dL$ along fermionic tangent vectors
vanishes. Computing this contraction we find that it does vanish provided the
rheonomic conditions (\ref{cond1})-(\ref{cond3}) hold (\cite{Castellani},
p.685). Thus the action is invariant only on the ``partial shell" of the
rheonomic conditions, and this invariance does not require the propagation equations.

However, the \textit{closure} of the supersymmetry transformations does
require also the propagation equations  (\ref{prop}) to hold\footnote{this can
be understood by checking Bianchi identities: after enforcing rheonomic
constraints on the curvatures, Bianchi identities are not identities anymore,
and other conditions may arise for them to hold. These other conditions are
the propagation equations.} , and therefore the supersymmetry algebra closes
only on shell.

The situation is drastically different when auxiliary fields are available to
close the supersymmetry algebra off-shell. Then one finds that the fermionic
contractions of $dL$ vanish identically without requiring any condition. This
can be checked for example in the so-called new minimal $D=4$, $N=1$
supergravity (or Sohnius-West model \cite{SohniusWest}), where the
superPoincar\'e algebra is enlarged and auxiliary fields (a 1-form and a
2-form) enter the game. In fact in this case the natural algebraic framework
is that of free differential algebras \cite{Castellani}, a generalization of
Lie algebras, whose dual formulation in terms of Cartan-Maurer equations is
generalized to contain also $p$-form fields.

\section{Ectoplasmic Integration with Integral Forms}

We would like to put in relation the so-called Ectoplasmic technique (Ethereal
Integration Theorem) with integral forms. The main point is to prove, by using
the integral forms, the so-called ``ectoplasmic integration theorem". This
theorem states that, given a function ${L}$ of the superspace (also known as
superspace action) on a curved supermanifold $\mathcal{M}$ whose geometry is
described by the supervielbein $E_{M}^{A}$ (see eq. (\ref{inM})), its
integral
\begin{equation}
I_{\mathcal{M}}=\int_{\mathcal{M}}E {L}\, [d^{n}xd^{m}\theta]\label{ecto00}%
\end{equation}
where $E$ is the superdeterminant of $E_{M}^{A}$, is equal to the following
integral
\begin{equation}
I_{\mathcal{S}}=\int_{\mathcal{S}}e\mathcal{D}^{m} {L}\left.  {}\right\vert
_{\theta=0}d^{n}x\label{ecto01}%
\end{equation}
where $e$ is the determinant of the vielbein $e_{n}^{a}$ of the bosonic
submanifold $\mathcal{S}$ of $\mathcal{M}$ (more precisely, $\mathcal{S}$ is
identified with the bosonic submanifold of $\mathcal{M}$ obtained by setting
to zero the fermionic coordinates). The expression $\mathcal{D}^{m} {L}\left.
{}\right\vert _{\theta=0}$ denotes the action of a differential operator
${\mathcal{D}}^{m}$ on the function ${L}$ evaluated at $\theta=0$.
$\mathcal{D}^{m}$ is a symbol denoting a differential operator of order $m$ in
the super derivatives. The form of the differential operator is difficult to
compute by usual Berezin integration since one has to evaluate the
supervielbein $E_{M}^{A}$ (at all orders of the $\theta$-expansion), compute
its superdeterminant and finally expand the product $E {L}$. That procedure
leads to the form $\mathcal{D}^{m}{L}\left.  {}\right\vert _{\theta=0}$, where
${\mathcal{D}}^{m}$ is a combination of super derivatives, ordinary
derivatives and non-derivative terms and the coefficients depend upon
curvature, torsion and higher derivative supergravity tensors. The relation
between $I_{\mathcal{M}} $ and $I_{\mathcal{S}}$ is easy in the case of flat
superspace since there is no superdeterminant to be computed and all
supergravity tensors drop out.

In order to circumvent this problem, Gates \textit{et al.} proposed a new
method to evaluate $\mathcal{D}^{m}{L}\left.  {}\right\vert _{\theta=0}$.
First, one has to select a closed superform (that we will denote by $L
^{(n|0)}$) with degree equal to the dimension of the bosonic submanifold. The
form must be closed on the complete supermanifold, namely $dL ^{(n|0)}=0,$
where $d$ is the differential on the full supermanifold. The closure of the
superform (and also its non-exactness) and the existence of a constant tensor
imply that a given component of $L ^{(n|0)}$ can be written in terms of this
tensor times an arbitrary function $\Omega(x,\theta)$ on the supermanifold.
All other components of $L ^{(n|0)}$ are either vanishing or written as
combination of derivatives of the arbitrary function $\Omega(x,\theta)$. The
coefficients of those combinations are related again to supergravity tensors.
The total result $L ^{(n|0)}$ is a superform whose coefficients are given in
terms of $\Omega(x,\theta)$, a combination of derivatives and supergravity
fields. The \textbf{Ethereal conjecture} is that the unknown function
$\Omega(x,\theta)$ coincides with the superspace action ${L}$ evaluated at
$\theta=0$.

The first step is to translate the definitions given by Gates \textit{et al.}
in term of integral forms. Then, we show that the integrals of eq.
(\ref{ecto00}) and eq. (\ref{ecto01}) can be viewed as integrals of integral
forms that can be related via the Poincar\'e dual. Finally, by changing the
Poincar\'e dual by a different embedding of the bosonic submanifold into the
supermanifold, we are able to show that indeed the function $\Omega(x,
\theta)$ does coincide with the superspace action ${L}$.

\subsection{From Ectoplasm to Integral Forms}

The integral of $L^{(n|0)}$ (which we will denote in the following with
$\omega^{(n|0)}$) on the bosonic submanifold $I_{\mathcal{S}}$ is defined as
follows
\begin{equation}
{I_{\mathcal{S}}=\int_{\mathcal{S}}i^{\ast}\omega\equiv\int_{\mathcal{M}^{m}%
}\hat{\omega}^{(n|0)}\big|_{\theta=0}}\label{ectoI}%
\end{equation}
where $\mathcal{S}\equiv\mathcal{M}^{n}\subset\mathcal{M}^{(n|m)}%
\equiv\mathcal{M}$ is the bosonic submanifold (obtained by setting to zero the
anticommuting variables in the transition functions) and $\hat{\omega}%
^{(n|0)}\big|_{\theta=0}$ is obtained from $\omega^{(n|0)}$ by setting to zero
both the dependence on $\theta$ and on 1-forms $d\theta$
\begin{equation}
{i^{\ast}\omega=\hat{\omega}^{(n|0)}\big|_{\theta=0}=\omega_{\lbrack
i_{1}\dots i_{n}]}(x,0)dx^{i_{1}}{\wedge}\dots{}{\wedge}dx^{i_{n}}%
}\label{sett}%
\end{equation}
Notice that this superform can be integrated on the bosonic submanifold being
a genuine $n$-form, and if the manifold $\mathcal{S}$ is curved we get
\begin{equation}
I_{\mathcal{S}}=\int_{\mathcal{S}}e\epsilon^{a_{1}\dots a_{n}}\omega_{\lbrack
a_{1}\dots a_{n}]}(x,0)\label{csA}%
\end{equation}
where we have denoted by Latin letters $a_{1},\dots,a_{n}$ the flat indices
and $e$ is the determinant of the vielbein $e_{i}^{a}$.

The first crucial observation is that $I_{S}$ can be also rewritten, following
the prescription described in sec. 2, as follows
\begin{equation}
{I_{\mathcal{S}}=\int_{\mathcal{M}^{(n|m)}}\omega\wedge\eta_{S}=\int
_{\mathcal{M}^{(n|m)}}\omega^{(n|0)}\wedge\theta^{m}\delta^{m}(d\theta
)}\label{ectoII}%
\end{equation}
where, as usual, we denote by $\theta^{m}$ the product of all fermionic
coordinates $\theta^{\alpha}$ and by $\delta^{m}(d\theta)$
%\begin{equation*}
%\delta ^{m}(d\theta )=\epsilon _{\alpha _{1}\dots \alpha _{m}}\delta
%(d\theta ^{\alpha _{1}}) \dots  \delta (d\theta ^{\alpha _{m}})
%\end{equation*}%
the wedge product of all Dirac delta functions $\delta(d\theta^{\alpha})$.
Then, the Poincar\'{e} dual in this case is $\eta_{S}=\theta^{m}\delta
^{m}(d\theta)$ which is the product of \textquotedblleft picture changing
operators" embedding the bosonic submanifold $\mathcal{S}$ into the
supermanifold $\mathcal{M}$ in the simplest way $\theta^{1}=\theta^{2} =
\dots=0$.

The integration is performed over the entire supermanifold. A simple
computation leads to the original result (\ref{ectoI}). This is clear since
integrating over the $d\theta$ has the effect that all components of
$\omega^{(n|0)}$ in the $d\theta$ directions are set to zero, leading to
$\hat{\omega}^{(n|0)}$. The Berezin integral over the coordinates $\theta$ is
simplified since the presence of the product $\theta^{m}$ forces us to pick up
the first component of $\hat{\omega}^{(n|0)}$, namely $\hat{\omega}%
^{(n|0)}|_{\theta=0}$ leading to the integral.

\subsection{Closure and Susy}

The important point about (\ref{ectoI}) is the invariance under supersymmetry.
The variation under supersymmetry of $\hat{\omega}^{(n|0)}$ is given by a
local translation in superspace
\begin{equation}
{\Delta_{\epsilon}\Big(\hat{\omega}^{(n|0)}\big|_{\theta=0}\Big)=(\Delta
_{\epsilon} \omega^{(n|0)})\big|_{\theta=0}=\epsilon^{\alpha}\Big({\frac
{\partial}{\partial\theta^{\alpha}}}+(\gamma^{i}\theta)_{\alpha}\partial
_{i}\Big)\hat{\omega}^{(n|0)}\big|_{\theta=0}=\epsilon^{\alpha}{\frac
{\partial}{\partial\theta^{\alpha}}}\hat{\omega}^{(n|0)}\big|_{\theta=0}%
}\label{susy}%
\end{equation}
where the first equality is due to the variation of the field components in
the expression of $\hat{\omega}^{(n|0)}$ (and therefore it does not matter
whether it is computed at $\theta=0$), the second equality is just the
expression of a susy transformation as a supertranslation in superspace. The
last term can be rewritten as follows:
\begin{equation}
{(\partial_{\alpha}\hat{\omega}_{i_{1}\dots i_{n}})dx^{i_{1}}\wedge\dots\wedge
dx^{i_{n}}=- n \, \partial_{\lbrack i_{1}}\omega_{i_{2}\dots i_{n}]\alpha}dx^{i_{1}%
}\wedge\dots\wedge dx^{i_{n}}}\label{last}%
\end{equation}
where we have used the closure of the superform $\omega^{(n|0)}=\omega
_{M_{1}\dots M_{n+1}}^{(n|0)}\,dZ^{M_{1}}\wedge\dots\wedge dZ^{M_{n+1}}$, (recall
(\ref{sett})) which implies
\begin{equation}
{\partial_{\lbrack M_{1}}\omega_{M_{2}\dots M_{n+1})}=0}\label{clos}%
\end{equation}
where the superindices $M_{1},\dots,M_{n+1}$ are graded-symmetrized. In this
way, the r.h.s. of (\ref{susy}) is a derivative w.r.t. to bosonic coordinates
$x^{i}$ and therefore, by integrating over $\mathcal{M}^{n}$, the integral
$I_{\mathcal{S}}$ in (\ref{ectoI}) vanishes. So, the key requirement to
guarantee the supersymmetric invariance of $I_{\mathcal{S}}$ is the
\textbf{closure} of $\omega^{(n|0)}$ as a superform in the full superspace.

Using (\ref{ectoII}), we observe that the integral form $\omega^{(n|0)}%
\wedge\theta^{m}\delta^{m}(d\theta)$ belongs to $\Omega^{(n|m)}$, namely the
space of top forms. The closure of $\omega^{(n|0)}$ implies the closure of
this integral form, since
\[
d\Big(\theta^{m}\delta^{m}(d\theta)\Big)=0.
\]
We also notice that if $\omega^{(n|0)}$ belongs to the $d$-cohomology
$H^{*}(\Omega^{(n|0)})$, so does the integral form, since $\theta^{m}%
\delta^{m}(d\theta)$ is in the $d$-cohomology $H^{*}(\Omega^{(0|m)})$ (which
are the class of forms with zero form degree and highest picture number, see
\cite{Catenacci:2010cs}). However, the converse is not true:
\begin{equation}
{d\Big(\omega^{(n|0)}\theta^{m}\delta^{m}(d\theta)\Big)=0\,\,\,\Rightarrow
\,\,\,d\omega^{(n|0)}=f_{\alpha}\theta^{\alpha}+g_{\alpha}d\theta^{\alpha}%
\,,}\label{conv}%
\end{equation}
$d\omega^{(n|0)}$ cannot be proportional to $\delta(d\theta)$ since it must be
a picture-zero form and $f_{\alpha}$ must belong to $\Omega^{(n|0)}$ while
$g_{\alpha}$ to $\Omega^{(n-1|0)}$. However, by consistency we have
$d\Big(f_{\alpha}\theta^{\alpha}+g_{\alpha}d\theta^{\alpha}\Big)=0$, which
implies that $df_{\alpha}=0$ and $f_{\alpha}=-dg_{\alpha}$. This yields
$f_{\alpha}\theta^{\alpha}+g_{\alpha}d\theta^{\alpha}=-d(g_{\alpha}%
\theta^{\alpha})$ which can be reabsorbed into a redefinition of
$\omega^{(n|0)}$, leading to a closed form.

Again we can check the susy invariance of $I_{S}$ in the form (\ref{ectoII}).
Performing the susy transformations leads to
\begin{equation}
{\Delta_{\epsilon}\Big(\omega^{(n|0)}\theta^{m}\delta^{m}(d\theta
)\Big)=\Big(\Delta_{\epsilon}\omega^{(n|0)}\Big)\theta^{m}\delta^{m}%
(d\theta)+\omega^{(n|0)}m(\epsilon\theta^{m-1})\delta^{m}(d\theta
)}\label{susyII}%
\end{equation}
where $\Delta_{\epsilon}\theta^{\alpha}=\epsilon^{\alpha}$ and $(\epsilon
\theta^{m-1})\equiv\epsilon_{\alpha_{1}\dots\alpha_{m}}\epsilon^{\alpha_{1}%
}\theta^{\alpha_{2}}\dots\theta^{\alpha_{m}}$. Due to the closure of
$\omega^{(n|0)}$ we are in the same situation as above: the partial derivative
w.r.t. to $\theta$ can be re-expressed as an $x$-derivative and its integral
is then zero. The second piece is zero because we integrate over $\theta$
\textit{\`{a} la} Berezin and, since $\omega^{(n|0)}$ is computed at
$\theta=0$ (being multiplied by $\theta^{m}$), the integral vanishes.

\subsection{Density Projection Operator}

Now, we need to understand the integral obtained in (\ref{ectoI}) in terms of
the superform $\omega^{(n|0)}$ by using the closure of it. We adopt the
description given by Gates and we follow the same derivation.

Let us now compute the expression in (\ref{ectoI}), namely we compute
$\hat{\omega}^{(n|0)}$ by passing to non-holonomic coordinates as follows
\begin{equation}
{\omega_{M_{1}\dots M_{n}}dZ^{M_{1}} {\wedge}\dots{\wedge}dZ^{M_{n}}%
=\omega_{\Sigma_{1}\dots\Sigma_{n}}E^{\Sigma_{1}}\wedge\dots{}\wedge
E^{\Sigma_{n}}\longrightarrow}\label{nonhol}%
\end{equation}%
\[
\hat{\omega}^{(n|0)}\big|_{\theta=0}=\Big(\omega_{\Sigma_{1}\dots\Sigma_{n}%
}E_{i_{1}}^{\Sigma_{1}}\dots E_{i_{n}}^{\Sigma_{n}}\Big)\big|_{\theta
=0}\epsilon^{i_{1}\dots i_{n}}d^{n}x
\]
where we denote by $\Sigma$ the non-holonomic super indices. So, we have:
\begin{equation}
{\hat{\omega}^{(n|0)}\big|_{\theta=0}=\Big(\omega_{I_{1}\dots I_{n}}E_{i_{1}%
}^{I_{1}}\dots E_{i_{n}}^{I_{n}}+\dots+\omega_{A_{1}\dots A_{n}}E_{i_{1}%
}^{A_{1}}\dots E_{i_{n}}^{A_{n}}\Big)\big|_{\theta=0}\epsilon^{i_{1}\dots
i_{n}}d^{n}x}\label{super}%
\end{equation}
and $E_{i}^{I}\big|_{\theta=0}=e_{i}^{I}$ is the bosonic vielbein of the
bosonic manifold $\ \mathcal{M}^{n}$ while $E_{i}^{A}\big|_{\theta=0}=\psi
_{i}^{A}$ where $\psi_{i}^{A}$ is the gravitino field of the supergravity
model underlying it. Then,
\begin{equation}
{\hat{\omega}^{(n|0)}\big|_{\theta=0}=\Big(\omega_{I_{1}\dots I_{n}}e_{i_{1}%
}^{I_{1}}\dots e_{i_{n}}^{I_{n}}+\dots+\omega_{A_{1}\dots A_{n}}\psi_{i_{1}%
}^{A_{1}}\dots\psi_{i_{n}}^{A_{n}}\Big)\big|_{\theta=0}\epsilon^{i_{1}\dots
i_{n}}d^{n}x}\label{superII}%
\end{equation}
Using $\psi_{I}^{A}e_{i}^{I}=\psi_{i}^{A}$, it yields
\begin{equation}
{\hat{\omega}^{(n|0)}\big|_{\theta=0}=e\Big(\omega_{I_{1}\dots I_{n}}%
+\dots+\omega_{A_{1}\dots A_{n}}\psi_{I_{1}}^{A_{1}}\dots\psi_{I_{n}}^{A_{n}%
}\Big)\big|_{\theta=0}\epsilon^{I_{1}\dots I_{n}}d^{n}x}\label{superIII}%
\end{equation}

The requirement that the superform must be closed, $d\omega^{(n|0)}=0$,
expressed in terms of the non-holonomic basis, implies that
\begin{equation}
{D_{[\Sigma_{1}}\omega_{\Sigma_{2}\dots\Sigma_{n+1})}^{(n|0)}+T_{~~[\Sigma
_{1}\Sigma_{2}}^{\Gamma}\omega_{|\Gamma|\Sigma_{3}\dots\Sigma_{n+1})}%
^{(n|0)}=0\,,}\label{eqnon}%
\end{equation}
%(the notation $[\Sigma_1 \dots \Sigma_n)$ denotes the graded symmetrisation of the indices)
where $T_{[\Sigma_{1}\Sigma_{2})}^{\Gamma}$ are the components of the torsion
computed in the non-holonomic basis. The form $\omega^{(n|0)}$ is defined up
to gauge transformations
\begin{equation}
{\delta\omega_{\lbrack\Sigma_{1}\dots\Sigma_{n+1})}^{(n|0)}=D_{[\Sigma_{1}%
}\Lambda_{\Sigma_{2}\dots\Sigma_{n+1})}+T_{~~[\Sigma_{1}\Sigma_{2}}^{\Gamma
}\Lambda_{|\Gamma|\Sigma_{3}\dots\Sigma_{n+1})}\,,}\label{gautr}%
\end{equation}
the notation $|\Gamma|$ excludes the index $\Gamma$ from the graded symmetrization.

The coefficients of the torsion satisfy the Bianchi identities
\begin{equation}
\label{bianchi}{\ D_{[\Sigma_{1}} T^{\Gamma}_{~\Sigma_{2} \Sigma_{3})} +
T_{(\Sigma_{1} \Sigma_{2}}^{\Lambda}\, T^{\Gamma}_{~~|\Lambda| \Sigma_{3}]} =
R^{\Gamma}_{~~[\Sigma_{1} \Sigma_{2} \Sigma_{3})} }%
\end{equation}
where $R^{\Gamma}_{[\Sigma_{1} \Sigma_{2} \Sigma_{3})}$ are the components of
the curvature. We also recall that $[D_{\Sigma}, D_{\Gamma}] = T_{~~\Sigma
\Gamma}^{\Lambda} D_{\Lambda}+ R^{I}_{~J, \Sigma\Gamma} M_{I}^{~J}$ where
$M_{I}^{~J}$ are the Lorentz generators.

The Bianchi identities become non-trivial equations when some of the
components $\omega_{\Sigma_{1}\dots\Sigma_{n}}$ are set to a given value. For
example by choosing some of $\omega_{I_{1}\dots I_{p}A_{p+1}\dots A_{n}}$ with
spinorial indices equal to zero, such that the next one
\[
\omega_{I_{1}\dots I_{p-{1}}A_{p}\dots A_{n}}=\Omega(x,\theta)\,t_{I_{1}\dots
I_{p-{1}}A_{p}\dots A_{n}}%
\]
can be set equal a constant tensor $t_{I_{1}\dots I_{p-{1}}A_{p}\dots A_{n}}$
(combination of Dirac gamma matrices and invariant tensors) where
$\Omega(x,\theta)$ is a superfield. The other components can be fixed by
solving the Bianchi identities and it is easy to show that
\[
\omega_{I_{1}\dots I_{n}}=f_{I_{1}\dots I_{n}}^{A_{1}\dots A_{m}}D_{A_{1}%
}\dots D_{A_{m}}\Omega+\dots
\]
where the dots are other tensors constructed out of curvature and derivative
of it. The coefficients $f_{I_{1}\dots I_{n}}^{A_{1}\dots A_{m}}$ are
combinations of constant tensors. Inserting the solution of the Bianchi
identities into the cumulative expression (\ref{superIII}) one gets an
expression of the density projector $\mathcal{D}^{m}$
\[
{\hat{\omega}^{(n|0)}\big|_{\theta=0}=e\Big(f_{I_{1}\dots I_{n}}^{A_{1}\dots
A_{m}}D_{A_{1}}\dots D_{A_{m}}\Omega+\dots+ f_{A_{1} \dots A_{n}} \psi_{I_{1}%
}^{A_{1}}\dots\psi_{I_{n}}^{A_{n}}\Omega\Big)\big|_{\theta=0}\epsilon
^{I_{1}\dots I_{n}}d^{n}x}%
\]%
\begin{equation}
\label{denpro}\equiv e\mathcal{D}^{m}\Omega\big|_{\theta=0}d^{n}x
\end{equation}
The exponent $m$ denotes the maximal number of spinorial derivative. This
conclude this review part on the density projection operator and we are
finally in position to present a proof of the theorem.

\subsection{Proof of the Ectoplasmic Integration Theorem}

At this point we need to study the other side of the Ectoplasmic Integration
Theorem, namely we have to describe the integral $I_{\mathcal{M}}$ in terms of
a superform. For that we recall eq. (\ref{inM}): the integral of a top
integral form $\omega^{(n|m)}$ reads
\[
I_{\mathcal{M}} \equiv \int_{\mathcal{M}^{(n|m)}}\omega^{(n|m)} =
\]%
\[
=
\int_{\mathcal{M}^{(n|m)}}%
\omega_{\lbrack I_{1}\dots I_{n}][A_{1}\dots A_{m}]}^{(n|m)}(x,\theta
)E^{I_{1}}\wedge\dots{}\wedge E^{I_{n}{\ }}\wedge\delta(E^{A_{1}})\wedge
\dots\wedge{}\delta(E^{A_{m}}) =%
\]%
\[
=\epsilon^{I_{1}\dots I_{n}}\epsilon^{A_{1}\dots A_{m}}\int_{\mathcal{M}%
^{(n|m)}}E\omega_{\lbrack I_{1}\dots I_{n}][A_{1}\dots A_{m}]}^{(n|m)}%
(x,\theta)d^{n}x\delta^{m}(d\theta)=
\]%
\begin{equation}
=\int E\,\epsilon^{I_{1}\dots I_{n}}\epsilon^{A_{1}\dots A_{m}}\omega_{\lbrack
I_{1}\dots I_{n}][A_{1}\dots A_{m}]}^{(n|m)}(x,\theta) %
\label{ectoIII}%
\end{equation}
where in the last step we have stripped out the integration over the 1-forms
$d^{n}x$ and over the Dirac delta's $\delta^{m}(d\theta)$. We are then left
with the integral over the bosonic coordinates and the Berezin integral over
the fermionic coordinates. The latter can be performed by taking the
derivatives with respect to the fermionic coordinates of the product $E
\omega$ where $\omega=\epsilon^{I_{1}\dots I_{n}}\epsilon^{A_{1}\dots A_{m}%
}\omega_{\lbrack I_{1}\dots I_{n}][A_{1}\dots A_{m}]}^{(n|m)}(x,\theta)$.

The main point here is the following: the superfield $\Omega(x,\theta)$
appearing in the expression of $\hat{\omega}^{(n|0)}$ -- obtained by
\textquotedblleft integrating\textquotedblright\ the Bianchi identities with
some constraints -- has apparently nothing to do with the superfield
$\epsilon^{A_{1}\dots A_{m}}\omega_{\lbrack I_{1}\dots I_{n}][A_{1}\dots
A_{m}]}^{(n|m)}(x,\theta)$ appearing in (\ref{ectoIII}). Thus, to prove the
ectoplasmic integration formula one has to verify that they indeed coincide.
In order to do that we observe that the superform $\hat{\omega}^{(n|0)}$
belongs to the space $\Omega^{(n|0)}$ which has vanishing picture number.
Thus, in order to integrate it we need to change its picture by inserting
Picture Changing Operators of the form
\begin{equation}
\label{PCOb}Y=M^{A}(x,\theta)\delta(\psi^{A})+N^{A}(x,\theta,dx)\delta
^{\prime}(\psi^{A})+\dots
\end{equation}
where $\psi^{A}$ are the gravitino superfields (also denoted by $E^{A}$ in the
present work) and the dots stand for terms with higher derivative of Dirac
delta functions. The functions $M^{A}(x,\theta)$, $N^{A}(x,\theta,dx),\dots$
are needed to impose $dY=0$.

Therefore, we can construct a top integral form from $\hat{\omega}^{(n|0)}$
as
\begin{equation}
{\omega^{(n|m)}=\hat{\omega}^{(n|0)}\epsilon_{A_{1}\dots A_{m}}\Phi^{A_{1}%
}\dots\Phi^{A_{m}}\delta^{m}(\psi)\,,}\label{newINT}%
\end{equation}
which has the correct picture number and the correct form number to be
integrated on $\mathcal{M}^{(n|m)}$. The symbol $\Phi^{A_i}$ denotes
a function of $\theta$'s such that $d \Phi^{A} \delta(\psi^A) =0$, giving
a new arbitrary expression for the picture changing operator.
Then it is easy to show that, by
integrating by parts (using the superderivative appearing in the coefficients
of $\hat{\omega}^{(n|0)}$),  the integral form obtained is proportional to the
superfield $\Omega(x,\theta),$ and being the integral top-form sections of a
one-dimensional line bundle (the Berezinian), we conclude that the superfields
appearing in the expansion of $\hat{\omega}^{(n|0)}$ and of $\epsilon
^{A_{1}\dots A_{m}}\omega_{\lbrack I_{1}\dots I_{n}][A_{1}\dots A_{m}%
]}^{(n|m)}(x,\theta)$ are simply proportional and they can be chosen to be the same.

In other words, this corresponds to modifying the picture changing operators,
but remaining in the same cohomology class. That implies that the two
integrals are indeed equal since the delta functions appearing in $Y$ soak up
the gravitinos appearing in the density projection operator $\mathcal{D}^{m}$.

Let summarize the main steps of the proof. We start by showing that both the
integral $I_{\mathcal{M}}$ and $I_{\mathcal{S}}$ can be written in terms of integral
forms. The former is viewed as an integral of a density which is the
coefficient of a top form of $\Omega^{(n|m)}$. The second integral
$I_{\mathcal{S}}$ is converted into an integral of an integral form by
introducing a suitable picture changing operator $\theta^{m}\delta^{m}%
(d\theta)$. However, the choice of the picture changing operator is arbitrary
and therefore it can be changed into the new form (\ref{PCOb}), such that
the gravitons $\psi^A$ appear as arguments of the delta functions.
 Finally, the
computation of the integral $I_{\mathcal{S}}$ projects out all components of
the combination except a superfield $\Omega(x,\theta)$ which can be chosen to
be equal to the density of $I_{\mathcal{M}}$.

%%%%%%%%%%%%%%%%%%%%%%%%%%%%%%%%%%%%%%%%%

\section*{Acknowledgments}

We thank L. Andrianopoli, R. D'Auria and M. Trigiante for useful discussions
concerning integral forms and their use in supergravity theories.

%%%%%%%%%%%%%%%%%%%%%%%%%%%%%%%%%%%%%%%%%%

\section{Appendix A.}

We do not wish here to give an exhaustive and rigorous treatment of integral
forms. A systematic exposition of the matter can be found in the references
quoted in Section 2 .

As we said in section 2, the problem is that we can build the space
$\Omega^{k}$ of $k$-superforms out of basic 1-superforms $d\theta^{i}$ and
$dx^{i}$ and their wedge products, however the products between the
$d\theta^{i}$ are necessarily commutative, since the $\theta^{i}$'s are odd
variables. Therefore, together with a differential operator $d$, the spaces
$\Omega^{k} $ form a differential complex
\begin{equation}
0\overset{d}{\longrightarrow}\Omega^{0}\overset{d}{\longrightarrow}\Omega
^{1}\dots\overset{d}{\longrightarrow}\Omega^{n}\overset{d}{\longrightarrow
}\dots\label{comA}%
\end{equation}
which is bounded from below, but not from above. In particular there is no
notion of a top form to be integrated on the superspace $\mathbb{R}^{p|q}$.

The space of integral forms is obtained by adding to the usual space of
superforms a new set of basic ``forms" $\delta(d\theta)$, together with the
derivatives $\delta^{(p)}(d\theta)$, (derivatives of $\delta(d\theta)$ must be
introduced for studying the behaviour of the symbol $\delta(d\theta)$ under
sheaf morphisms i.e. coordinate changes, see below) that satisfies certain
formal properties.

These properties are motivated and can be deduced from the following heuristic approach.

In analogy with usual distributions acting on the space of smooth functions,
we think of $\delta(d\theta)$ as an operator acting on the space of superforms
as the usual Dirac's delta ``measure" (more appropriately one should refer to
the theory of de Rham's currents \cite{griffith-harris}), but this matter will
not be pursued further). We can write this as
\[
\left\langle f(d\theta),\delta(d\theta)\right\rangle =f(0),
\]
where $f$ is a superform. This means that $\delta(d\theta)$ kills all
monomials in the superform $f$ which contain the term $d\theta$. The
derivatives $\delta^{(n)}(d\theta)$ satisfy
\[
\left\langle f(d\theta),\delta^{(n)}(d\theta)\right\rangle =-\left\langle
f^{\prime}(d\theta),\delta^{(n-1)}(d\theta)\right\rangle =(-1)^{n}f^{(n)}(0),
\]
like the derivatives of the usual Dirac $\delta$ measure.

Moreover we can consider objects such as $g(d\theta)\delta(d\theta)$, which
act by first multiplying by $g$ then applying $\delta(d\theta)$ (in analogy
with a measure of type $g(x)\delta(x)$), and so on. The wedge products (when
defined, note that we cannot in general multiply distributions of the same
coordinates) among these objects satisfy some simple relations such as (we
will omit the symbol $\wedge$ of the wedge product):
\[
dx^{I}dx^{J}=-dx^{J}dx^{I}\,,\quad dx^{I}d\theta^{j}=d\theta^{j}dx^{I}\,,
\]%
\[
d\theta^{i}d\theta^{j}=d\theta^{j}d\theta^{i}\,,\quad\delta(d\theta
)\delta(d\theta^{\prime})=-\delta(d\theta^{\prime})\delta(d\theta),
\]%
\[
d\theta\delta(d\theta)=0\,,\quad d\theta\delta^{\prime}(d\theta)=-\delta
(d\theta).
\]

The most noticeable relation is the unfamiliar minus sign appearing in
$\delta(d\theta)\delta(d\theta^{\prime})=-\delta(d\theta^{\prime}%
)\delta(d\theta)$ (indeed this is natural if we interpret the delta ``forms"
as de Rham's currents) but can be also easily deduced from the above heuristic
approach. To prove this formula we recall the usual transformation property of
the usual Dirac's delta function
\[
\delta(ax+by)\delta(cx+dy)=\frac{1}{\left\vert \det\left(
\begin{array}
[c]{cc}%
a & b\\
c & d
\end{array}
\right)  \right\vert }\delta(x)\delta(y)
\]
for $x,y\in\mathbb{R}$. We note now that in the case under consideration the
absolute value must be dropped in the formula above, because the scaling
properties of $\delta(d\theta)$ are driven by $\int\delta\left(
d\theta\right)  [d\left(  d\theta\right)  ]=1.$ Under a rescaling
$\theta\rightarrow\lambda\theta$ ( $\lambda\in\mathbb{R}$ ) we must have
$\int\delta\left(  \lambda d\theta\right)  [d\left(  \lambda d\theta\right)
]=1,$ but now $d\theta$ is bosonic and hence the fermionic $d\left(
d\theta\right)  $ scales as $d\left(  \lambda d\theta\right)  =\lambda
d\left(  d\theta\right)  $. Hence setting $a=0,b=1,c=1$ and $d=1$, in the
correct formula:%
\[
\delta(ad\theta+bd\theta^{\prime})\delta(cd\theta+dd\theta^{\prime})=\frac
{1}{\det\left(
\begin{array}
[c]{cc}%
a & b\\
c & d
\end{array}
\right)  }\delta(d\theta)\delta(d\theta^{\prime})
\]
the anticommutation property of Dirac's delta function of $d\theta$'s follows.

An interesting and important consequence of this procedure is the existence of
\textit{negative degree forms}, which are those that by multiplication reduce
the degree of a forms (e.g. $\delta^{\prime}(d\theta)$ has degree $-1$).

We introduce also the \textit{picture number} by counting the number of delta
functions (and their derivatives) and we denote by $\Omega^{r|s}$ the space of
$r$-forms with picture $s$. For example, in the case of $\mathbf{R}^{p|q}$,
the integral form
\[
dx^{[K_{1}}\dots dx^{K_{l}]}d\theta^{(i_{l+1}}\dots d\theta^{i_{r})}%
\delta(d\theta^{\lbrack i_{r+1}})\dots\delta(d\theta^{i_{r+s}]})\
\]
is an $r$-from with picture $s$. All indices $K_{i}$ are antisymmetrized,
while the first $r-l$ indices are symmetrized and the last $s$ are
antisymmetrized. By adding derivatives of delta forms
$\delta^{(p)}(d\theta)$, even negative form-degree can be considered, e.g. a
form of the type:
\[
\delta^{(n_{1})}(d\theta^{i_{1}})\dots\delta^{(n_{s})}(d\theta^{i_{s}})
\]
is a $-(n_{1}+\dots + n_{s})$-form with picture $s$. Clearly $\Omega^{k|0}$ is
just the space $\Omega^{k}$ of superforms, for $k\geq0$.

Integral forms form a new complex as follows
\begin{equation}
\dots\overset{d}{\longrightarrow}\Omega^{(r|q)}\overset{d}{\longrightarrow
}\Omega^{(r+1|q)}\dots\overset{d}{\longrightarrow}\Omega^{(p|q)}\overset
{d}{\longrightarrow}0
\end{equation}
We now briefly discuss how these forms behave under change of coordinates,
i.e. under sheaf morphisms. For generic morphisms it is necessary to work with
infinite formal sums in $\Omega^{r|s}$ as the following example clearly shows.

Suppose $(\tilde{\theta}^{1})=\theta^{1}+\theta^{2}$ , $(\tilde{\theta}%
^{2})=\theta^{2}$ be the odd part of a morphism. We want to compute
\[
(\delta\left(  d\tilde{\theta}^{1}\right)  )=\delta\left(  d\theta^{1}%
+d\theta^{2}\right)
\]
in terms of the above forms. We can formally expand in series about, for
example, $d\theta^{1}:$
\[
\delta\left(  d\theta^{1}+d\theta^{2}\right)  =\sum_{j} {\frac{\left(
d\theta^{2}\right)  ^{j} }{j!}}\delta^{(j)}(d\theta^{1})
\]
Recall that any usual superform is a polynomial in the $d\theta,$ therefore
only a finite number of terms really matter in the above sum, when we apply it
to a superform. Indeed, applying the formulae above, we have for example,
\[
\left\langle (d\theta^{1})^{k},\sum_{j} {\frac{\left(  d\theta^{2}\right)
^{j} }{j!}}\delta^{(j)}(d\theta^{1})\right\rangle =(-1)^{k}(d\theta^{2})^{k}%
\]
Notice that this is equivalent to the effect of replacing $d\theta^{1}$ with
$-d\theta^{2}.$ We could have also interchanged the role of $\theta^{1}$ and
$\theta^{2}$ and the result would be to replace $d\theta^{2}$ with
$-d\theta^{1}.$ Both procedures correspond precisely to the action we expect
when we apply the $\delta\left(  d\theta^{1}+d\theta^{2}\right)  $ Dirac
measure. We will not enter into more detailed treatment of other types of morphisms.

\eject
\vfill

%%%%%%%%%%%%%%%%%%%%%%%%%%%%%%%%%%%%%%%%%%%%%%%%%%%%%%%%%%%%%%%%%%

\end{document}